\documentclass[sigconf]{acmart}
\AtBeginDocument{%
  \providecommand\BibTeX{{%
    \normalfont B\kern-0.5em{\scshape i\kern-0.25em b}\kern-0.8em\TeX}}}

\setcopyright{none}
\copyrightyear{To appear in proceedings of CHI 2021}
\acmYear{2021}
\acmDOI{10.1145/3411764.3445176}

\acmPrice{}
\acmISBN{978-1-4503-8096-6/21/05}

\usepackage{float}
\usepackage{multirow}
\usepackage[skip=5pt]{caption}

\begin{document}

\title{Does Interaction Improve Bayesian Reasoning with Visualization?}

\author{Ab Mosca}
\affiliation{ \institution{Tufts University}}
\email{amosca01@cs.tufts.edu}

\author{Alvitta Ottley}
\affiliation{ \institution{Washington University in St. Louis}}
\email{alvitta@wustl.edu}

\author{Remco Chang}
\affiliation{ \institution{Tufts University}}
\email{remco@cs.tufts.edu}

\renewcommand{\shortauthors}{Mosca, et al.}

\begin{abstract}
  Interaction enables users to navigate large amounts of data effectively, supports cognitive processing, and increases data representation methods.  
However, there have been few attempts to empirically demonstrate whether adding interaction to a static visualization improves its function beyond popular beliefs.    
In this paper, we address this gap. 
We use a classic Bayesian reasoning task as a testbed for evaluating whether allowing users to interact with a static visualization can improve their reasoning.  
Through two crowdsourced studies, we show that adding interaction to a static Bayesian reasoning visualization does not improve participants' accuracy on a Bayesian reasoning task. In some cases, it can significantly detract from it. Moreover, we demonstrate that underlying visualization design modulates performance and that people with high versus low spatial ability respond differently to different interaction techniques and underlying base visualizations.
Our work suggests that interaction is not as unambiguously good as we often believe; a well designed static visualization can be as, if not more, effective than an interactive one.

\end{abstract}




\maketitle

\section{Introduction} 

Interaction is core to visualization design and is a vital mode of communication between the user and visual system. In the visualization community, the study of interaction ranges from defining interaction~\cite{dimara2020What, yi2007Toward, heer2012Interactive}, to understanding the interplay between interaction and cognition~\cite{liu2010Mental, pohl2012User}, to leveraging user interactions to improve analytics~\cite{brown2012Disfunction, endert2012Semantic}. However, investigations into the value of adding interaction to a static design are rare and results are varied.
 
In some cases, the value-add of interaction to visualization is clear. In \textit{The Value of Visualization} van Wijk explains ``interaction is generally considered as good", and argues that it is invaluable to tasks such as allowing users to explore more data than can fit on a screen, and to customizing new visualization methods~\cite{wijk2005Value}. Heer and Shneiderman echo this sentiment in their taxonomy of interactive dynamics for visual analysis~\cite{heer2012Interactive}. 
Additionally, it has been argued that interaction is valuable due to its ability to amplify or illustrate user cognition~\cite{yi2007Toward, pohl2012User, liu2010Mental}. A recent study by Zhi et al. found that adding interaction to a storytelling visualization increased engagement~\cite{zhi2019linking}. Studies of multimedia instruction have shown that interactivity can increase deep learning and learning transfer~\cite{evans2007Interactivity, wang2011Impact}. 
      
However, the literature does not uniformly support interaction as an indisputable means of improving visualization. In fact, in \textit{The Value of Visualization} immediately after expressing the ``good" aspects of interaction van Wijk states that ``one could advocate the opposite: interaction should be avoided,'' and explains that interaction can negatively impact visualization by increasing subjectivity, and the user's perceptual and exploration costs~\cite{wijk2005Value}. Lam designed a framework that accounts for potential costs of interaction in information visualization, and encourages designers to weigh the cost against potential gains~\cite{lam2008Framework}. A study by Theis et al.~\cite{theis2016Ergonomic} comparing task performance on interactive and static uncertainty visualizations found no significant difference in error rate between the two. And a study by Ragan et al.~\cite{ragan2012Spatial} comparing outcomes of a pictorial learning activity given an interactive or automatic view control found no significant differences between the two.       

In this paper, we investigate the following research question: ``What value can interaction add to a static Bayesian reasoning visualization?'' We use a Bayesian reasoning task, because it is a well defined but difficult reasoning problem, with a clear-cut correct answer~\cite{ottley2016Bayesian,micallef2012Assessing,khan2015Benefits}. Moreover, Bayesian reasoning can be summarized quite succinctly by conditional probabilities and Bayes rule, however this often fails to adequately communicate the real world situation represented by these numbers~\cite{gigerenzer1995How}. As a result, there has been a plethora of research on communicating Bayesian reasoning through static visualization~\cite{brase2009Pictorial, garcia2013Visual, kellenFacilitating2007, ottley2012Visually, tsai2011Interactive, friederichs2014Using, sedlmeier2001Teaching, spiegelhalter2011Visualizing, gigerenzer1995How, cole1989Understanding, cole1989Graphic, khan2018Interactive, bocherer2019How, ottley2016Bayesian, micallef2012Assessing, khan2015Benefits}.   

In addition to being an open problem area, communicating Bayesian reasoning is an ideal test bed for interaction because interaction is not imperative to the effectiveness of a Bayesian reasoning visualization like it is for most visual analytic systems, which are built to analyze large amounts of data. Static Bayesian reasoning visualizations typically do not represent more data than can fit on one screen. Thus, adding interaction does not add any otherwise obscured information to the visualization, it simply highlights or draws connections between information already present. This allows us to to isolate the value-add of interaction independent of data exploration and sensemaking. 

Based on prior work~\cite{ottley2016Bayesian, tsai2011Interactive, micallef2012Assessing, khan2015Benefits, khan2018Interactive}, we postulate that interaction can facilitate visual Bayesian reasoning, but its effects are modulated by: (1) the interaction technique, and (2) the visualization design. Moreover, we expect to see different effects for people with high versus low spatial ability~\cite{ottley2016Bayesian}. In this work, we aim to gain a better understanding of how these factors affect the value-add of interaction to a static Bayesian reasoning visualization. 

To this end, we run two Amazon Mechanical Turk studies. Experiment 1 investigates the effect of adding interactive checkboxes to three different static (or base) visualizations, which range in their use of Gestalt principles to effectively depict sub-populations of interest in a Bayesian reasoning task. Analysis of Experiment 1 shows that adding interactive checkboxes to a static Bayesian reasoning visualization does not significantly impact participants' reasoning accuracy. Moreover, we do not find a case in which interaction significantly improves participants' performance. 
Experiment 2 investigates the same three base visualizations, and expands the number of interaction techniques tested (two types of checkboxes, drag and drop, hover, and tooltips). In our analysis of Experiment 2, we again do not find any cases in which interaction significantly improves participants' performance on Bayesian reasoning. Moreover, we find that for participants with high spatial ability, \textit{hover} significantly decreases performance.     

To summarize, we make the following contributions: 
\begin{enumerate}
	\item We demonstrate that adding interaction to a static Bayesian reasoning visualization can (under certain circumstances) decrease users' accuracy on a Bayesian reasoning task.
	
	\item We provide empirical and observational evidence that the value-add of interaction to a static Bayesian reasoning visualization is dependant on two factors: design of the static visualization, and interaction technique. 
	
	\item We show that adding interaction to a static Bayesian reasoning visualization can lower accuracy of people with high spatial ability on a Bayesian reasoning task, and generally does not effect accuracy of people with low spatial ability on a Bayesian reasoning task. 
\end{enumerate}
\section{Related Work}

\textit{Interaction} and \textit{Bayesian Reasoning} are widely studied areas in visualization, but they are typically studied in isolation from each other. This paper focuses on the intersection of these two areas. By using Bayesian reasoning as a test bed for interaction techniques, we add to the body of knowledge on interactivity, and Bayesian reasoning visualizations. The following sections discuss related work studying the value add of interaction, Bayesian reasoning visualizations, and the intersection of the two.

\subsection{Value add of interaction}
Recent work by Dimara and Perin~\cite{dimara2020What} defines interaction (in visualization) as ``the interplay between a person and a data interface involving a data-related intent, at least one action from the person and an interface reaction that is perceived as such.'' Similarly, Yi et al.~\cite{yi2007Toward} and Heer and Shneiderman~\cite{heer2012Interactive} construct taxonomies of interactions for visualization. Others have endeavored to better explain why interaction is useful to visualization from a cognitive processing standpoint~\cite{liu2010Mental, pohl2012User}. In addition to theorizing and categorizing interaction, work has been done designing novel interaction techniques, for example \cite{carpendale2012Beyond, goffin2020Interaction, lee2012Beyond, wybrow2014Interaction}, and identifying how visualization designers can leverage interaction to learn about users and create customized visualizations \cite{brown2012Disfunction, endert2012Semantic}. 

The majority of work on defining, categorizing, theorizing, and leveraging interaction for visualization focuses on visual analytic systems built to help users explore large amounts of data. The value-add of interaction in such cases is relatively clear; actions such as panning, zooming, and selecting subsets are indisputably essential to exploring datasets too large to reasonably fit on a single screen~\cite{wijk2005Value, heer2012Interactive}. Although there are potential costs to interaction~\cite{lam2008Framework, wijk2005Value}, in the case of visual analytic systems, these are often outweighed by benefits. Furthermore, in visual analytic systems interaction is often seen as an essential support for users' cognitive processing; it is viewed as the embodiment of sensemaking and knowledge discovery~\cite{yi2007Toward, pohl2012User, liu2010Mental, pike2009science}.  
In contrast, the value-add of interaction to static visualizations is not clear cut. Here we are specifically referring to interactions that do not reveal otherwise hidden data; i.e. they \textit{do not add} any information to the visualization. This consideration is vital because the reasoning problems that we consider in this paper are notoriously difficult and studies suggest that adding interaction to a challenging task can result in cognitive overload~\cite{mayer2001Cognitive}.

There is a sampling of prior work that are relevant to the investigations in this paper. 
For example, Zhi et al.~\cite{zhi2019linking} found that adding interaction through brushing and linking to a storytelling visualization increased engagement. 
Theis et al.~\cite{theis2016Ergonomic} compared static and interactive versions of an uncertainty data visualization and concluded based on accuracy and speed that the static visualization was preferable to its interactive counterpart.
Ragan et al.~\cite{ragan2012Spatial} compared comprehension and detail recall in a pictorial learning activity across interactive versus automatic view controls, and found no significant differences between the two. 
Note that all of these studies include an A-B test between one static and one interactive visualization. Our work adds nuance to this body of work by investigating the value-add of interactivity with a multi-factor experimental design.  
  
\subsection{Bayesian Reasoning}
An area in which Bayesian reasoning problems are prevalent is medical decision making. The standard example of a Bayesian reasoning problem in this context is the mammography problem~\cite{gigerenzer1995How}: 
\vspace{-.5em}
\begin{quote}
    The probability of breast cancer is 1\% for women at age forty who participate in routine screening. If a woman has breast cancer, the probability is 80\% that she will get a positive mammography. If a woman does not have breast cancer, the probability is 9.6\% that she will also get a positive mammography.\\
    \textit{A woman in this age group had a positive mammography in a routine screening. What is the probability that she actually has breast cancer?}
\end{quote}
\vspace{-.5em}

Using conditional probabilities and Bayesian reasoning to solve this problem is difficult for patients and physicians alike~\cite{eddy1982Probabilistic}. Given the importance of accurate medical risk communication and understanding, numerous studies have investigated how to aid people in Bayesian reasoning. One common approach is to change the text from probability format to frequency format~\cite{gigerenzer1995How, tsai2011Interactive}.    
Another approach is visualization. 

Numerous studies have investigated the effect of visualization on Bayesian reasoning with a variety of different designs.
Different techniques tested include Euler diagrams~\cite{brase2009Pictorial, kellenFacilitating2007, micallef2012Assessing, khan2015Benefits}, frequency grids or icon arrays~\cite{garcia2013Visual, kellenFacilitating2007, micallef2012Assessing, ottley2012Visually, sedlmeier2001Teaching, khan2015Benefits, tsai2011Interactive, bocherer2019How}, decision trees~\cite{friederichs2014Using, sedlmeier2001Teaching, spiegelhalter2011Visualizing, khan2015Benefits, bocherer2019How}, ``beam cut" diagrams~\cite{gigerenzer1995How}, probability curves~\cite{cole1989Understanding}, contingency tables~\cite{cole1989Understanding, cole1989Graphic}, double trees~\cite{khan2015Benefits, khan2018Interactive, bocherer2019How}, flow charts~\cite{khan2015Benefits}, pipe diagrams~\cite{khan2015Benefits}, Sankey diagrams~\cite{khan2015Benefits}, and unit squares~\cite{bocherer2019How}. 

Despite all of these studies, there is still no clear consensus on the best visualization for Bayesian reasoning. 
Several studies compared multiple Bayesian reasoning visualizations to each other and to text~\cite{ottley2016Bayesian, micallef2012Assessing, khan2015Benefits}. All of these studies found that visualization did not significantly improve users' accuracy in performing a Bayesian reasoning task compared to text. Findings from Ottley et al.~and Micallef et al.~indicate that there may be a detrimental effect on users' ability to perform Bayesian reasoning when visualizations and numerical text descriptions are presented together~\cite{ottley2016Bayesian, micallef2012Assessing}. Ottley et al.~\cite{ottley2016Bayesian} shed light on a significant performance gap between people with high and low spatial ability on a Bayesian reasoning task, and indicated that optimal visualization and text designs for people with high spatial ability differ from those for people with low spatial ability.

To the best of our knowledge, only two studies have investigated the effect of interactive Bayesian reasoning visualizations, and the results are conflicting. Tsai et al.~\cite{tsai2011Interactive} tested a frequency grid with interactive checkboxes against textual descriptions of the Bayesian reasoning problem. They found the interactive frequency grid resulted in a significant increase in accuracy compared to a textual description of the problem with statistics in probability format. Khan et al.~\cite{khan2018Interactive} compared a static and interactive double tree diagram of the Bayesian reasoning problem. They added interaction via drag and drop, and found the interactive double tree diagram significantly decreased accuracy in performing the Bayesian reasoning task compared to static. Our goal is to provide context for these conflicting results. By identifying how static visualization design and users' spatial ability affect the value-add of interaction to a static visualization, we can shed light on confounding factors that may explain differences in prior results.
    
\section{Research Goals}
Given conflicting prior results on how interaction impacts Bayesian reasoning visualizations, the overarching goal of this paper is to empirically test whether adding interaction to a static Bayesian reasoning visualization 
can improve performance on a reasoning task. We hypothesize that the mixed results of prior work are partially due to confounding factors between experiments, such as underlying visualization designs. Additionally, based on work by Ottley et al.~\cite{ottley2016Bayesian} which demonstrates that spatial ability is a  significant predictor of one's accuracy on Bayesian inference, we expect to see different effects of adding interaction to a Bayesian reasoning visualization across different levels of spatial ability.  
Specifically, we investigate the following:

\begin{itemize}
	\item[] \textbf{RQ1}: Does adding interaction to a static visualization improve accuracy on a Bayesian reasoning task? 
	\item[] \textbf{RQ2}: Is the effect of interaction modulated by the effectiveness of the underlying static visualization? 
	\item[] \textbf{RQ3}: Do users with different levels of spatial ability react to an interactive Bayesian reasoning visualization differently? 
\end{itemize}

In the remainder of this paper we describe two crowdsourced experiments designed to begin answering these research questions. These experiments are a step towards a deeper understanding of interaction in visualization. By empirically demonstrating what specific factors lead to performance gains and losses when making a static visualization interactive, we hope to lay the ground work for better interactive visualization design with evidence-based design guidelines.

\begin{table*}[t]
\caption{Three interactive and static conditions used in Experiment 1. Full size images are available in supplementary materials.}
\begin{tabular}{ccccc}
 &  & \multicolumn{3}{c}{\textsc{Base Visualization}} \\ \toprule
 &  & \textit{grouped} & \textit{aligned} & \textit{randomized} \\ \midrule
\multirow{2}{*}{\rotatebox{90}{\textsc{Interaction}}} 
    & \rotatebox{90}{\textit{cbAll}} 
    & \includegraphics[width=.29\linewidth]{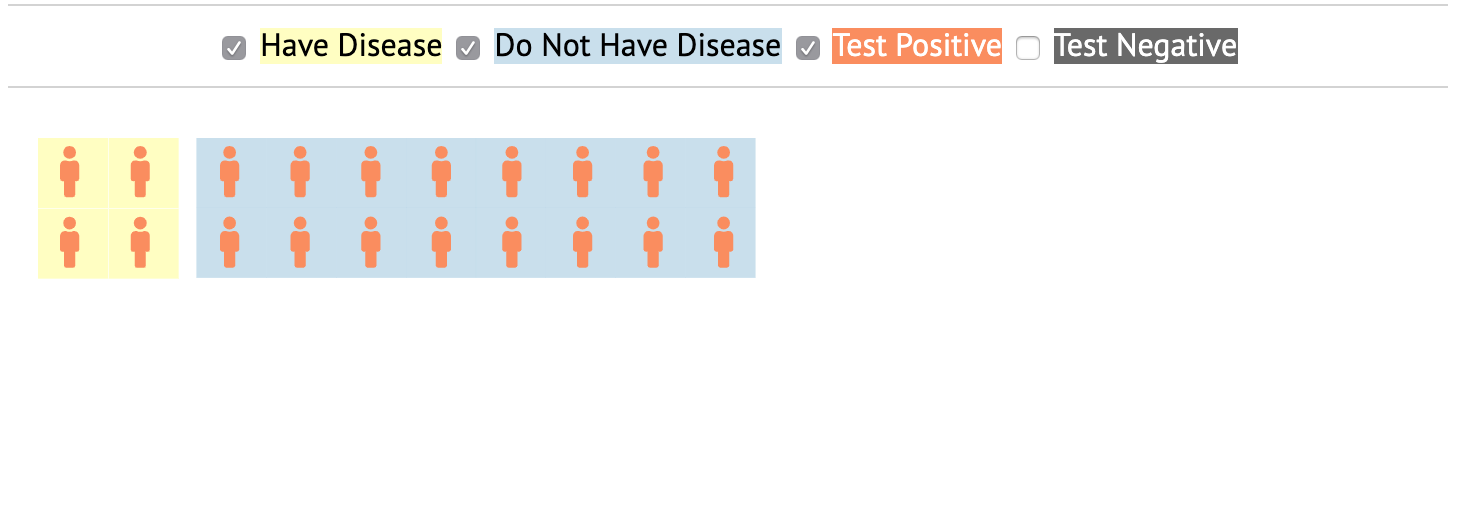}
    & \includegraphics[width=.29\linewidth]{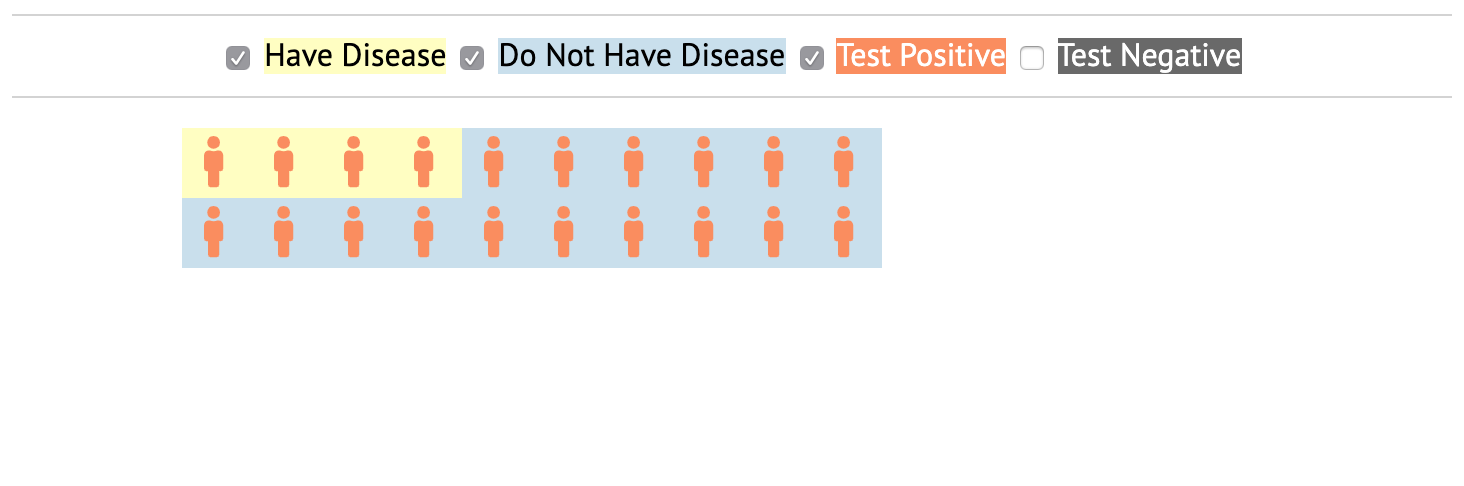}  
    & \includegraphics[width=.29\linewidth]{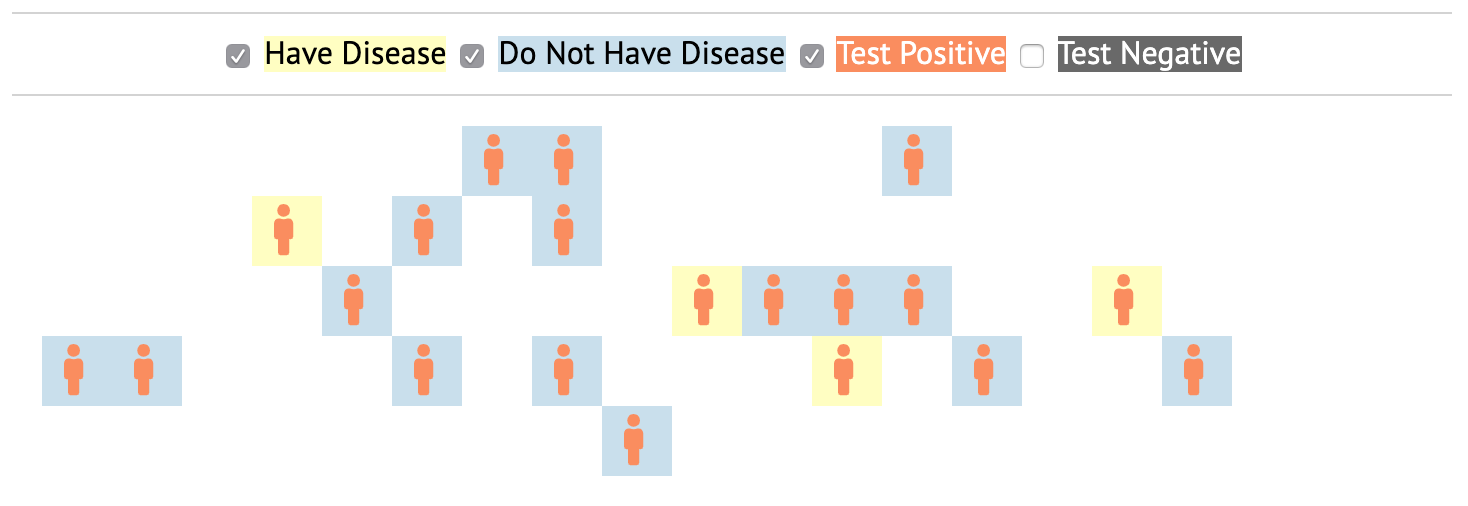}
    \\ 
    & \rotatebox{90}{\textit{static}}
    & \includegraphics[width=.29\linewidth]{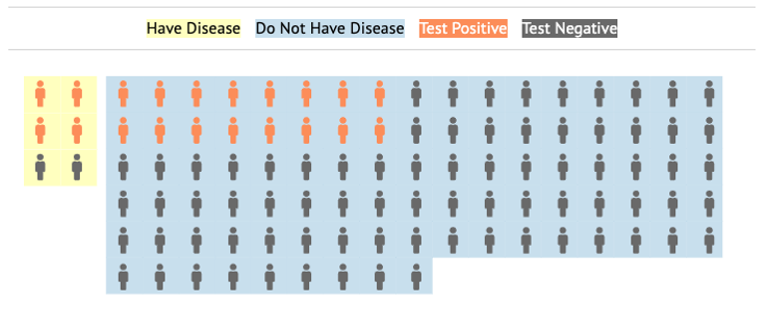}
    & \includegraphics[width=.29\linewidth]{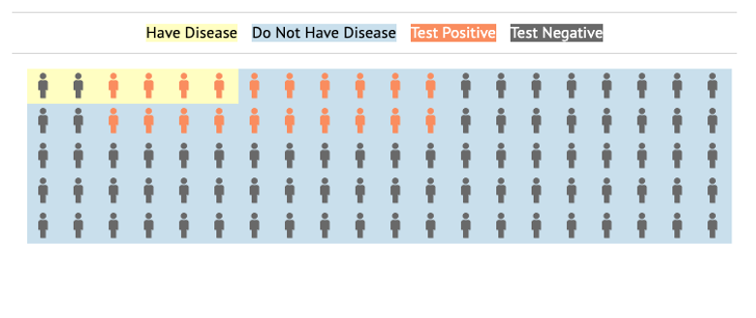}  
    & \includegraphics[width=.29\linewidth]{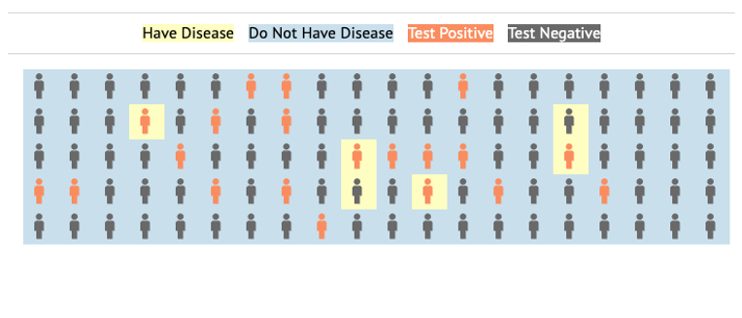} \\ \bottomrule
\end{tabular}
\label{tab:tabPilotVis}
\Description[Examples of stimuli for Exp. 1]{Examples of each base visualization design and the cbAll interaction (top row) and static version (bottom row). The cbAll visualizations show what the visualization would look like if everything except Test Negative is checked off.}
\end{table*}

\section{Experiment 1}
Although interaction is commonly used in data visualization, efforts to define \textit{interactivity} are ongoing~\cite{dimara2020What}. Additionally, there is no consensus on the best visualization for Bayesian reasoning~\cite{micallef2012Assessing, khan2015Benefits, ottley2016Bayesian}. As a result, the design space for interactive Bayesian reasoning visualizations is large. 
For this experiment, we simplify the visualization design space by focusing on variants of \textit{icon arrays} -- one of the most popular and well-studied visualizations in this context~\cite{micallef2012Assessing,ottley2016Bayesian,ottley2019Curious}. Guided by prior work on interactive Bayesian reasoning visualizations that found positive results~\cite{tsai2011Interactive}, we narrow the interaction design space to a single category, \textit{checkboxes}. 
In short, this experiment examines whether adding interactive \textit{checkboxes} to variations of static icon arrays improves accuracy in a Bayesian reasoning task.   

\subsection{Visualization Designs}
We present users with a Bayesian reasoning problem concerning a disease in the population and the false positive and negative rates associated with testing for the disease (Section \ref{sec:questions}). 
Each stimulus is an icon array that encodes the four key sub-populations of the problem: \textsc{Have Disease}, \textsc{Do Not Have Disease}, \textsc{Test Positive}, and \textsc{Test Negative}. 
Examples of interactive and static stimuli are shown in Table \ref{tab:tabPilotVis}. Below, we describe each experimental factor: 

\begin{itemize}
    \item \textbf{Base Visualization}: \{ \textit{grouped}, \textit{aligned}, \textit{randomized} \}
    \item \textbf{Interaction}: \{ \textit{checkboxes}, \textit{static} \}
\end{itemize}

\subsubsection{\textbf{Base Visualizations}}

We observe three primary designs of icon arrays in the literature which are loosely based on theories for how to facilitate Bayesian reasoning. For example, some researchers propose that representing randomness can more accurately communicate the inherent uncertainly in the problem space~\cite{han2011Representing}. Others hypothesize that spatially grouping visual elements aids reasoning~\cite{micallef2012Assessing}. Based on these theories, we design three variations for \textit{icon arrays} by changing the types of contextual placements of icons (see examples in Table~\ref{tab:tabPilotVis}).  
Following guidelines of Bertin~\cite{bertin}, background color was used to differentiate between members of the population who \textsc{Have Disease} versus \textsc{Do Not Have Disease}, and icon color was used to differentiate between members of the population who \textsc{Test Positive} versus \textsc{Test Negative}. The base visualizations differed in their use of Gestalt principles~\cite{gestalt} to perceptually group the sub-populations in the Bayesian reasoning problem. Each base visualization design is described in detail below: 

\begin{itemize}
	\item[] \textit{Grouped: }The \textit{grouped} icon array uses spatial grouping. It shows the sub-populations \textsc{Have Disease} and \textsc{Do Not Have Disease} in two separate grids of icons. Additionally, the \textsc{Test Positive} sub-population is in a block aligned at the top left of the visualization.
	This design is similar to the hybrid Euler-frequency grid diagram used by Micallef et al.~\cite{micallef2012Assessing}. 
	
	\item[] \textit{Aligned: }The \textit{aligned} icon array shows all icons in one 5 X 20 grid. It aligns the sub-population \textsc{Have Disease} in a block at the top left of the grid, and the sub-population \textsc{Test Positive} in a block at the top middle of the grid. A similar design was used by Brase et al.~\cite{brase2009Pictorial}, and Ottley et al.~\cite{ottley2016Bayesian,ottley2019Curious}.
	
	\item[] \textit{Randomized: }The \textit{randomized} icon array does not spatially group any sub-populations; icons representing members of each of the four  sub-populations are randomly distributed in a 5 X 20 grid. Similar designs are used in medical risk communication by Han et al.~\cite{han2011Representing}.
\end{itemize}

\subsubsection{\textbf{Interaction}}
We add \textit{checkboxes} to traditional icon arrays which allow the user to hide or show visual elements, and create a more explicit link between the text and the graphical encodings. We chose the checkbox interaction (1) for continuity with (and replication of) prior work ~\cite{tsai2011Interactive}, which found positive results from adding checkboxes to a Bayesian reasoning visualization; and (2) because the checkboxes enable participants to manipulate the visualization such that it directly encodes the answer to the Bayesian reasoning question without any additional distractors i.e. users can remove all information irrelevant to answering the questions posed.

As a default, when the interactive visualizations loaded on the page all checkboxes were checked, meaning the interactive visualizations initially look identical to the static version. When using the checkboxes, one of \{\textsc{Have Disease}, \textsc{Do Not Have Disease}\} as well as one of \{\textsc{Test Positive}, \textsc{Test Negative}\} had to be checked for any sub-populations to show on the visualization. The top row of Table \ref{tab:tabPilotVis} shows an example of what the visualization would show if everything except \textsc{Test Negative} were checked.  We call this interaction technique \textit{cbAll} (i.e. checkboxes where all boxes are initially checked).  

\subsection{Task}
\label{sec:questions}

We run a between-subjects 2 \{\textit{interaction}\} \textsc{x} 3 \{\textit{base visualization}\} factor experiment. In the study, participants are asked to answer a Bayesian reasoning problem given textual and visual representations. For continuity with prior work, the textual description and question components of each stimulus were consistent with those used by Ottley et al.~\cite{ottley2016Bayesian}:
 
\begin{itemize}
\item[] \textit{Textual description}: \\ There is a newly discovered disease, Disease X, which is transmitted by a bacterial infection found in the population. There is a test to detect whether or not a person has the disease, but it is not perfect. Here is some information about the current research on Disease X and efforts to test for the infection that causes it.  

There is a total of 100 people in the population. Out of the 100 people in the population, 6 people actually have the disease. Out of these 6 people, 4 will receive a positive test result and 2 will receive a negative test result. On the other hand, 94 people do not have the disease (i.e., they are perfectly healthy). Out of these 94 people, 16 will receive a positive test result and 78 will receive a negative test result.  
\item[] \textit{Questions}: \\
	(a) How many people will test positive? \_ \_ \_ \\
	(b) Of those who test positive, how many will actually have the disease? \_ \_ \_ 
\end{itemize}

\subsection{Participants}
We recruited 530 participants from Amazon Mechanical Turk. Participation was restricted to workers in the United States with an approval rating of greater than $90$ percent. Participants were paid a base rate of $\$1.80$ for participation plus a bonus of $\$0.10$ for every correct answer. 

Before analysis, participants who skipped entire sections of the experiment or did not follow instructions ($N = 3$), and participants who self-identified as colorblind  ($N = 55$) were dropped from the data set. 
This left $N = 472$ participants distributed among stimuli as shown in Table \ref{tab:pilotN}. Demographics of participants are shown in Table \ref{tab:pilotDemo}. 

\subsection{Procedure}
The experiment followed an approved protocol per Tufts University's IRB, and was posted as a HIT on Amazon Mechanical Turk. Workers who accepted the HIT followed a link to the experiment. After providing informed consent, participants were taken to an instruction page explaining the experiment.
This page demonstrated what the legend for a \textit{static} visualization would look like versus the \textit{cbAll} legend. After the instruction page, participants were shown one of the six experimental stimuli. Participants could take as much time with the stimulus as they wanted before clicking a button to view the questions to answer. 
After completing the main task, participants were asked to complete a short demographic questionnaire, the paper folding test (VZ-2) from Ekstrom, French, \& Hardon~\cite{paperFolding} to measure spatial ability\footnote{Due to space constraints, spatial ability analysis for Experiment 1 is included in supplemental materials.}, and to provide any additional feedback they wished. 

    
    \begin{table}[ht]
     \caption{Sample size (N) for each condition in Experiment 1.}
     \resizebox{0.4\textwidth}{!}{
     \centering
    \begin{tabular}{ccccc} \toprule
            & \textit{grouped} & \textit{aligned} & \textit{randomized} & Total \\ \midrule
    \textit{cbAll}   & 64     & 100     & 82   & 246 \\ 
    \textit{static}  & 86    & 70    & 70  &  226 \\ 
    Total & 150 & 170 & 152 & 472 \\ \bottomrule
    \end{tabular}}
    \label{tab:pilotN}
    \end{table}
    
  
    \begin{table}[ht]
    \caption{Experiment 1 participant demographics.}
    \resizebox{0.45\textwidth}{!}{
     \centering
    \begin{tabular}{ll} \toprule
    \textit{Measure} & \textit{Percentages} \\ \midrule
   
    N                                                                                               & 472                                                                                                                                                    \\ 
    Age                                                                                             & \begin{tabular}[c]{@{}l@{}}18-24: 6.4\%, 25-39: 64.4\%, 40-49: 17.2\%, \\ 50-59: 7.6\%, 60+: 4.4\%\end{tabular}                                        \\ 
    Gender                                                                                          & \begin{tabular}[c]{@{}l@{}}Female: 38.3\%, Male: 61.0\%, \\ Non-Binary: 0.8\%\end{tabular}                                                             \\ 
    Education                                                                                       & \begin{tabular}[c]{@{}l@{}}High School: 28.6\%, Bachelors: 56.6\%, \\ Masters: 10.2\%, PhD: 1.5\%, Other: 3.2\%\end{tabular}                           \\ 
    \begin{tabular}[c]{@{}l@{}}Expertise with \\ Statistical \\ Visualization\end{tabular}          & \begin{tabular}[c]{@{}l@{}}Novice: 21.4\%, Low-intermediate: 20.1\%, \\ Intermediate: 32.3\%, \\ High-intermediate: 17.4\%, Expert: 8.5\%\end{tabular} \\ 
    \begin{tabular}[c]{@{}l@{}}Statistical Training \\ 1(none) - \\ 5 (highly trained)\end{tabular} & \begin{tabular}[c]{@{}l@{}}1: 32.4\%, 2: 22.5\%, 3: 16.7\%,\\ 4: 15.3\%, 5: 12.5\%\end{tabular}                                                        \\ \bottomrule
    \end{tabular}}
    \label{tab:pilotDemo}
    \end{table}
    

\subsection{Research Questions}
We analyze data from Experiment 1 to answer these questions:  

\begin{itemize}
	\item[] \textbf{Q1.1} \textbf{Does adding interaction to a static visualization improve accuracy on Bayesian reasoning task?} We investigate whether participants who use an interactive visualization will be more accurate in answering the Bayesian reasoning question than participants who saw a static visualization.
	\item[] \textbf{Q1.2} \textbf{Is the effect of interaction modulated by the underlying static visualization design?}
	We investigate whether differences in accuracy will be modulated by the base visualization design (\textit{grouped, aligned, randomized}).
\end{itemize}

\subsection{Findings}
For analysis, participants' answers were considered correct only if they answered both parts of the two-part question correctly (this approach is consistent with prior work~\cite{ottley2016Bayesian, ottley2019Curious}). Our analysis script is included in supplemental materials, however under the guidelines of our IRB we are unable to release the data.

\subsubsection{\textbf{Does adding interaction to a static visualization improve accuracy on Bayesian reasoning task?}}
\label{sec:exp1_analysis}

\begin{figure}[ht]
    \includegraphics[width=0.45\textwidth]{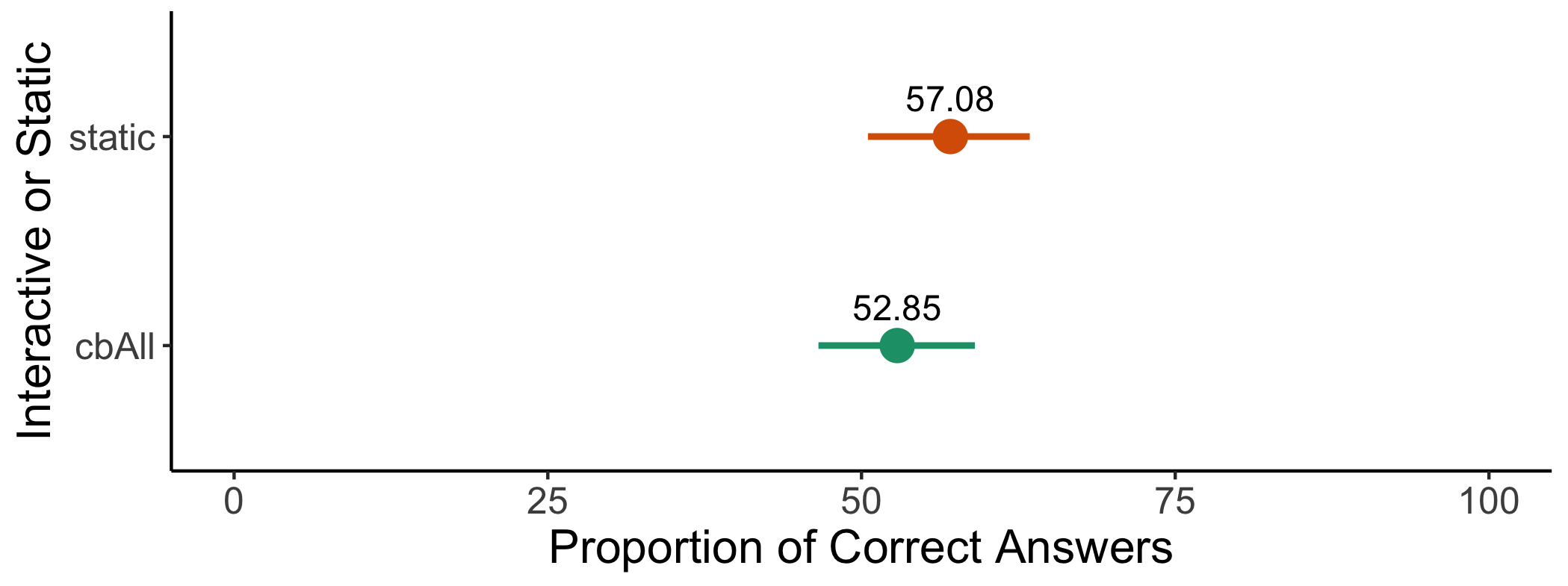}
    \caption{Proportion of participants answering the Bayesian reasoning task correctly given an interactive (\textit{cbAll}) versus \textit{static} visualization. Bars represent a 95\% logit transformed confidence interval.}
    \label{fig:exp1_static_vs_int}
    \Description[cbAll and static perform similarly.]{Dot chart showing proportions of correct answers for participants assigned to the static vs cbAll visualization. 95 percent logit transformed confidence intervals are also shown. Proportions are about equal between the two, and confidence intervals overlap significantly.}
    \end{figure}

Figure \ref{fig:exp1_static_vs_int} shows proportions of participants answering the Bayesian reasoning question correctly in the \textit{cbAll} and \textit{static} conditions. We observe $53\%$ of the participants in the \textit{cbAll} condition entered the correct answers, whereas the \textit{static} visualization had a $57\%$ correct response rate.
We perform a 2-sample test for equality of proportions of $accuracy \sim interactive\_or\_static$ with the null hypothesis that there is no difference in proportions of correct answers. We find no statistically significant difference in accuracy between participants using the \textit{cbAll} versus the \textit{static} visualization ($\chi^2(1, N = 472) = 0.69, p = 0.41$), and therefore fail to reject the null hypothesis. 
\textit{Ultimately, we found no evidence that adding interaction improves accuracy on a Bayesian reasoning task.}  

    
    \begin{figure}[ht]
    \includegraphics[width=0.45\textwidth]{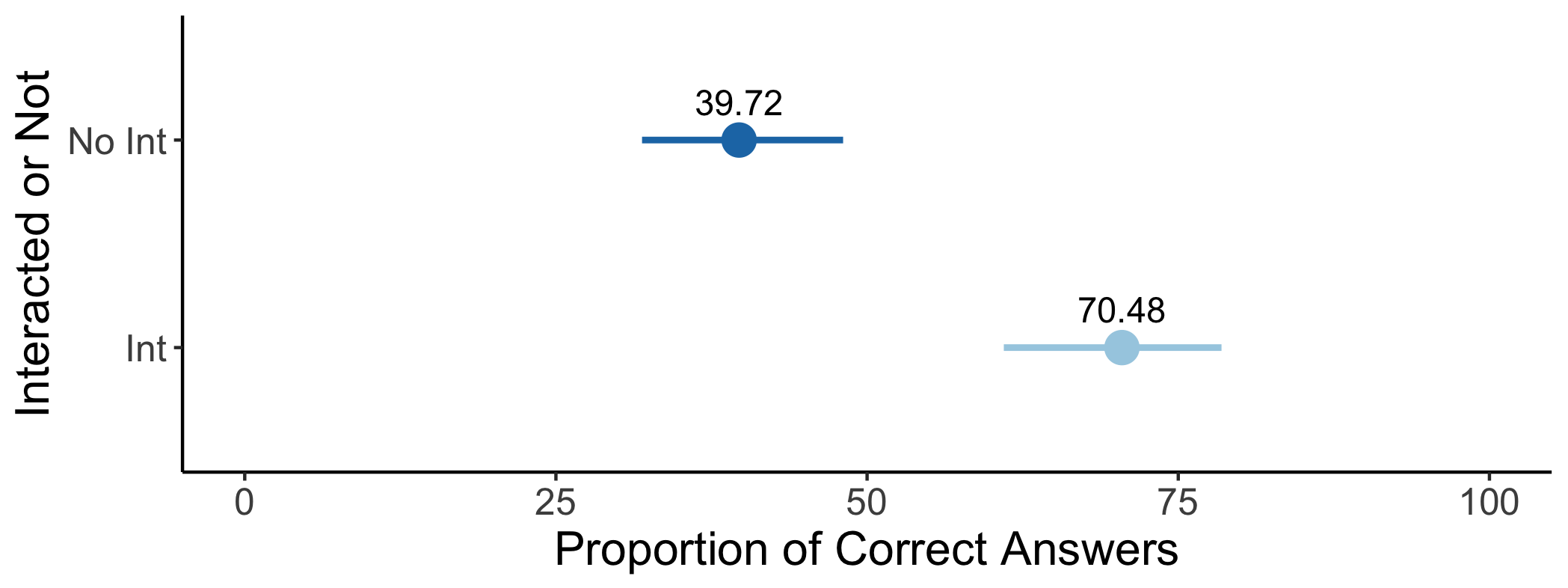}
    \caption{Proportion of participants assigned to \textit{cbAll} answering the Bayesian reasoning task correctly grouped by if they interacted or not. Bars represent a 95\% logit transformed confidence interval.}
    \label{fig:exp1_interacted_vs_did_not}
     \Description[Interacted vs did not interact perform differently.]{Dot chart showing proportions of correct answers for participants assigned to the cbAll visualization who did not interact with it vs those who did interact with it. 95 percent logit transformed confidence intervals are also shown. Proportions are significantly different (confidence intervals no not overlap at all). Participants who interacted significantly outperform those who did not.}
    \end{figure}
    

Next, we look at how many participants in the \textit{cbAll} condition interacted with the visualization they saw. Out of the $246$ participants assigned to \textit{cbAll}, only $43\%$ used the checkboxes on the visualization. Seventy percent of the participants who interacted with the visualization answered correctly, while only $40\%$ of those who did not interact answered correctly, as shown in Figure \ref{fig:exp1_interacted_vs_did_not}.
We perform a 2-sample test for equality of proportions of $accuracy \sim interacted\_or\_not$ with the null hypothesis that there is no difference in proportions of correct answers. We find a statistically significant difference in accuracy between participants who did and did not interact with the \textit{cbAll} visualization  ($\chi^2(1, N = 246) = 21.63, p < 0.001$), and therefore reject the null hypothesis. The 95\% confidence interval (using Wilson's score method) for the difference between the proportions is $[-43.48, -18.04]$. Taken together with the previous finding, this suggests that simply adding interaction does not guarantee use. However, \textit{the act of interacting may improve performance on a Bayesian reasoning task}. We note future work is needed to isolate if this is an artifact of participant engagement, or an effect of interacting.   


    
    \begin{figure}[ht]
    \includegraphics[width=0.45\textwidth]{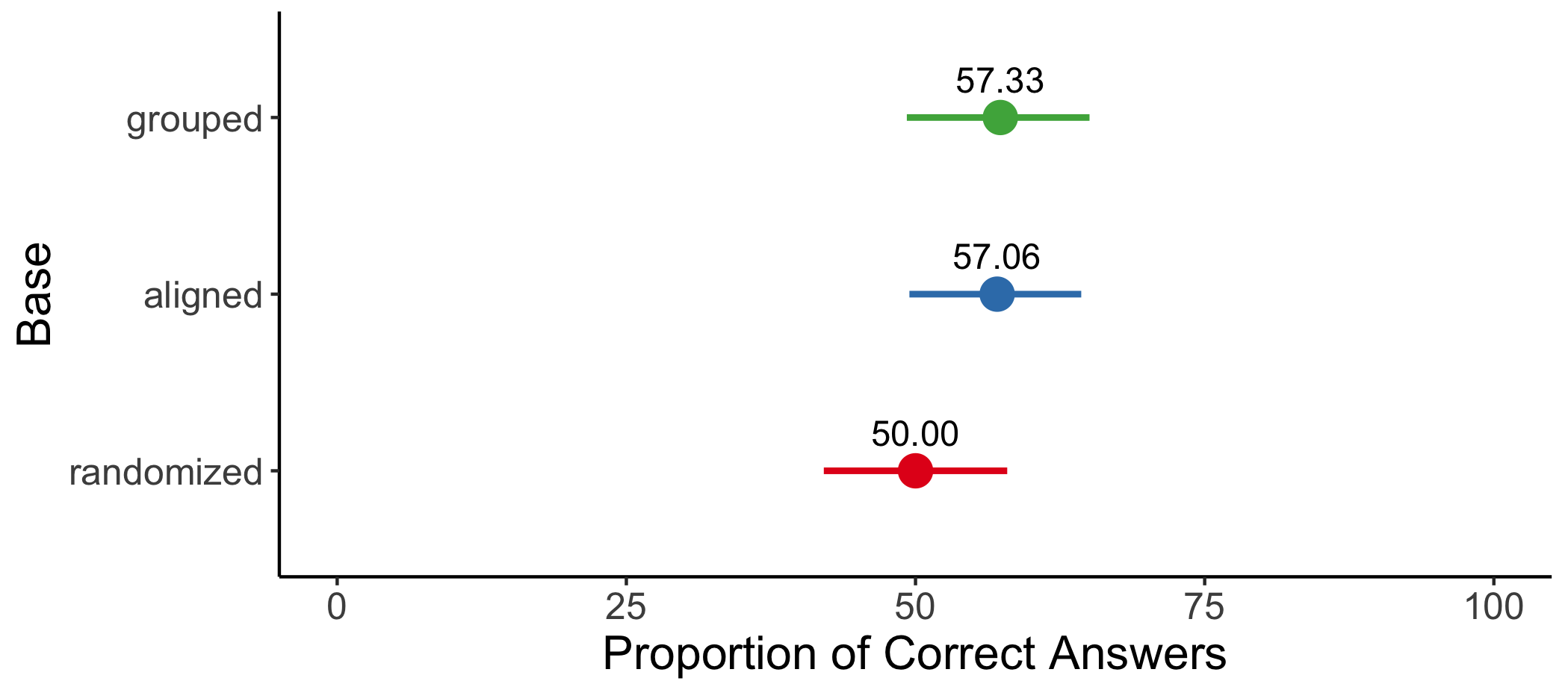}
    \caption{Proportion of participants answering the Bayesian reasoning task correctly by base visualization. Bars represent a 95\% logit transformed confidence interval.}
    \label{fig:exp1_bases}
     \Description[Base visualizations perform similarly.]{Dot chart showing proportions of correct answers for participants assigned to each base visualization. 95 percent logit transformed confidence intervals are also shown. Proportions are about equal between all three, and confidence intervals overlap significantly.}
    \end{figure}   
    
  
    \begin{figure}[ht]
    \includegraphics[width=0.45\textwidth]{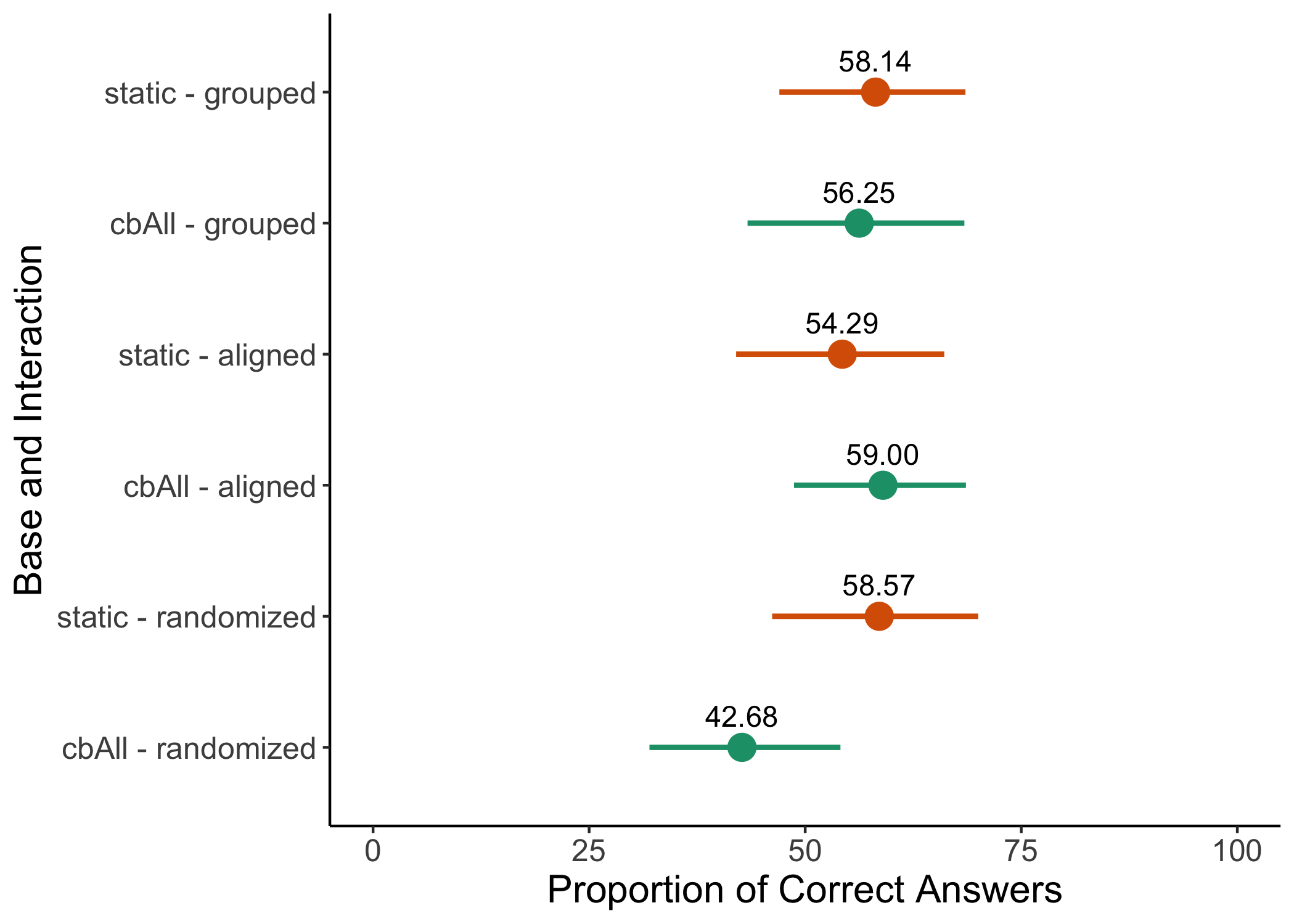}
    \caption{Proportion of participants answering the Bayesian reasoning task correctly by interaction and base visualization. Bars represent a 95\% logit transformed confidence interval.} 
    \label{fig:exp1_static_vs_int_by_base}
     \Description[cbAll - randomized visualization performs poorly.]{Dot chart showing proportions of correct answers for participants assigned to each combination of static or cbAll and base visualization. 95 percent logit transformed confidence intervals are also shown. Proportions are about equal in all cases, except for the cbAll - randomized visualization which performs noticeably (but not significantly) worse than the other visualizations.}
    \end{figure}
    

\subsubsection{\textbf{Is the effect of interaction modulated by the underlying static visualization design?}}

To investigate whether base visualization design has an effect on participants' accuracy we perform a 3-sample test for equality of proportions of $accuracy \sim base\_visualization$ with the null hypothesis that there is no difference in proportions of correct answers. We find no significant difference in accuracy by base visualization ($\chi^2(2, N = 472) = 2.15, p = 0.34$), and therefore fail to reject the null hypothesis. 
As shown in Figure~\ref{fig:exp1_bases}, we observe near-equal proportions of correct answers for the three base representations, suggesting that the \textit{variations in the design of icon arrays had no significant effect on accuracy.}

To investigate if there is an interaction effect between base visualization and whether a visualization is interactive or static we perform a 6-sample test for equality of proportions of $accuracy \sim \{base\_visualization\} X \{interactive\_or\_static\}$ with the null hypothesis that there are no differences in proportions of correct answers. We find no significant difference ($\chi^2(5, N = 472) = 6.42, p = 0.27$), and therefore fail to reject the null hypothesis. This suggests that \textit{the value-add of interaction may not be modulated by base visualization design.} 

Notably, Figure \ref{fig:exp1_static_vs_int_by_base} shows proportions of correct answers are nearly identical given all combinations of base visualization and interactive or static, but that the proportion of correct answers is smaller on average for participants assigned the \textit{cbAll} version of the \textit{randomized} visualization. While this difference is not statistically significant, we note that practically speaking it is important to consider; particularly in the case of medical risk decision making, where small improvements on this task can lead to more informed and autonomous medical decisions. 

\subsection{Discussion} 
\label{sec:exp1_discussion}
Although it is a common belief that interactivity adds value to visualizations, investigations into it merits can reveal essential insights about the pros, cons, or missed opportunities in interactive visualization design. In this experiment, we used Bayesian reasoning -- a problem that is notoriously challenging for the general population -- and showed that adding interactive checkboxes to a Bayesian reasoning visualization does not significantly improve reasoning accuracy. Moreover, in this experiment we observe cases where adding interaction decreases average performance on the Bayesian reasoning task. Though these differences are not statistically significant, the lack of a clear value add suggests future work should continue to investigate the potential costs and benefits of interaction in this setting.      

Our analyses suggest that there may not be a value-add to making a static visualization interactive. Moreover, our observational findings suggest that the effect of adding interactivity to a static visualization may depend on the design of the visualization itself. 
We used three variations of icon arrays based on theories for how to facilitate Bayesian reasoning: \textit{grouped}, \textit{aligned}, and \textit{randomized}. We observed nearly identical accuracy between the interactive and static versions of the \textit{grouped} and \textit{aligned} designs, and an insignificant, but practically relevant, decrease in accuracy for the interactive version of the \textit{randomized} design. We speculate that one rationale for this outcome is that the combination of a challenging Bayesian problem with randomness and interactivity may have induced an extraneously high cognitive load. 
Some experts caution that adding interactivity to a significantly complicated task can result in cognitive overload~\cite{mayer2001Cognitive}. Due to the lack of perceptual grouping, the \textit{randomized} base visualization induces more cognitive load than the \textit{grouped} and \textit{aligned} bases. Based on this and our observational findings, we postulate that adding interaction to the \textit{randomized} base may have caused cognitive overload in participants.    

An important observation is that a sizable portion of our study population assigned to the interactive visualization did not interact with it ($57\%$). There are a combination of factors that can explain this result.  
No interaction could be indicative of participants who were either confused about the task or were simply clicking through to get paid, as discussed in Section \ref{sec:exp1_analysis}. 
Alternatively, it is possible that participants did not want to interact. Existing work indicates that people may not engage with interactive visualizations as much as previously thought~\cite{boy2015storytelling} and there have been reports of media venues such as \textit{The New York Times}, scaling back their creation of interactive visualization in lieu of static images\footnote{\label{foot:nyt}Why We Are Doing Fewer Interactives (Archie Tse, The New York Times): https://github.com/archietse/malofiej-2016/blob/master/tse-malofiej-2016-slides.pdf}.
While understanding if there is a value-add of interaction is an important step to user-centered interactive visualization design, we recognize that understanding users' perceived value of interaction is also crucial. 

\section{Experiment 2}
The findings of Experiment 1 suggest that
adding a \textit{checkbox} interaction to a static Bayesian reasoning visualization has little to no effect on reasoning accuracy. Experiment 2 expands on this study by exploring the effect of different interaction techniques. Specifically, we compare the effects of adding \textit{two types of checkboxes}, \textit{drag and drop}, \textit{hover}, and \textit{tooltips} to the three \textit{icon array} base visualizations used in Experiment 1.  

\begin{table*}[t!]
\caption{Five interactive conditions used in Experiment 2. Full size images are available in supplementary materials.}
\begin{center}
\begin{tabular}{ccccc}
 &  & \multicolumn{3}{c}{\textsc{Base Visualization}} \\ \toprule
 &  & \textit{grouped} & \textit{aligned} & \textit{randomized} \\ \midrule
\multirow{4}{*}{\rotatebox{90}{\textsc{Interaction}}} 
    & \rotatebox{90}{\textit{cbAll} \textbackslash \textit{cbNone}} 
    & \includegraphics[width=.29\linewidth]{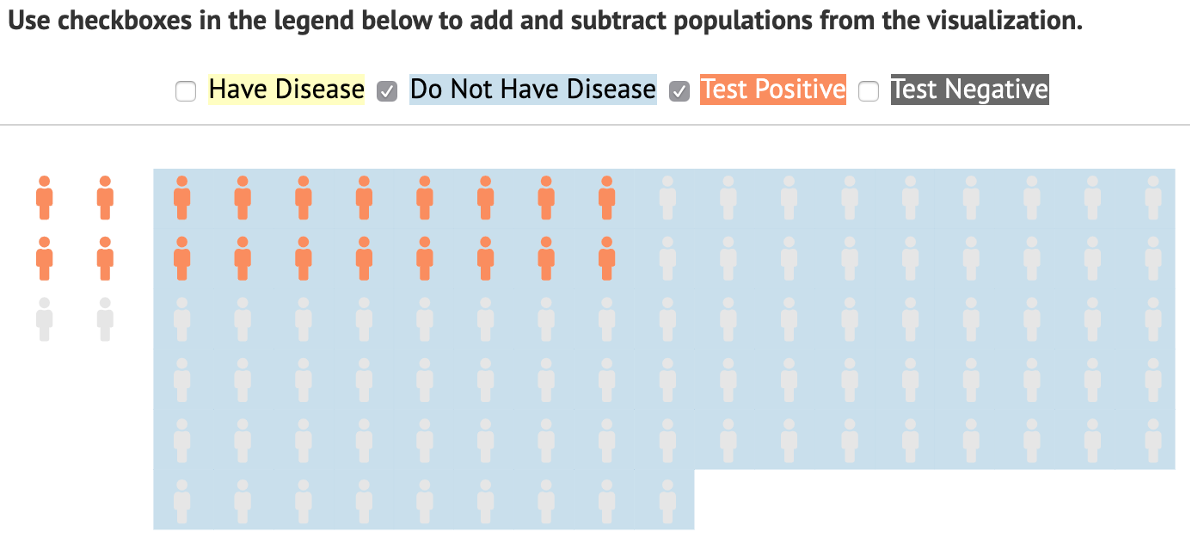}  
    & \includegraphics[width=.29\linewidth]{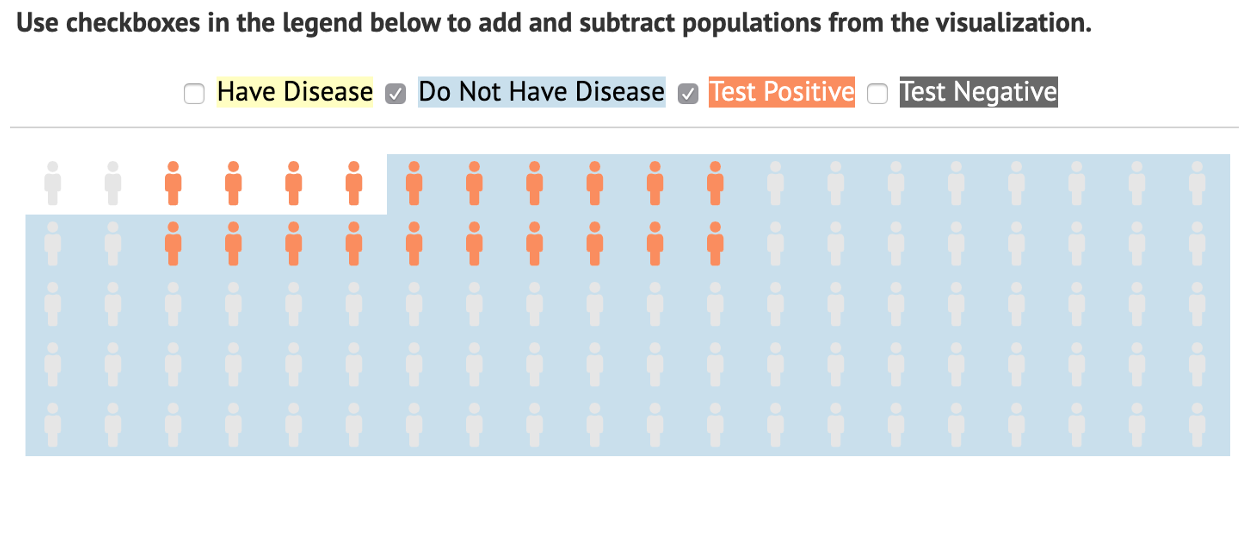}  
    & \includegraphics[width=.29\linewidth]{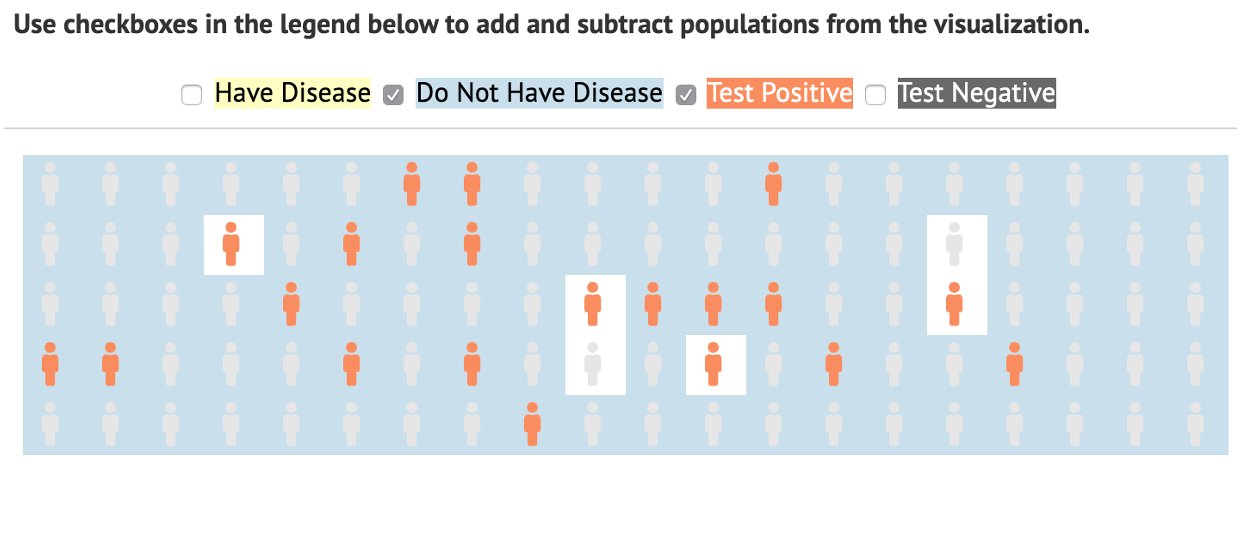}
    \\
    & \rotatebox{90}{\textit{drag}} 
    & \includegraphics[width=.29\linewidth]{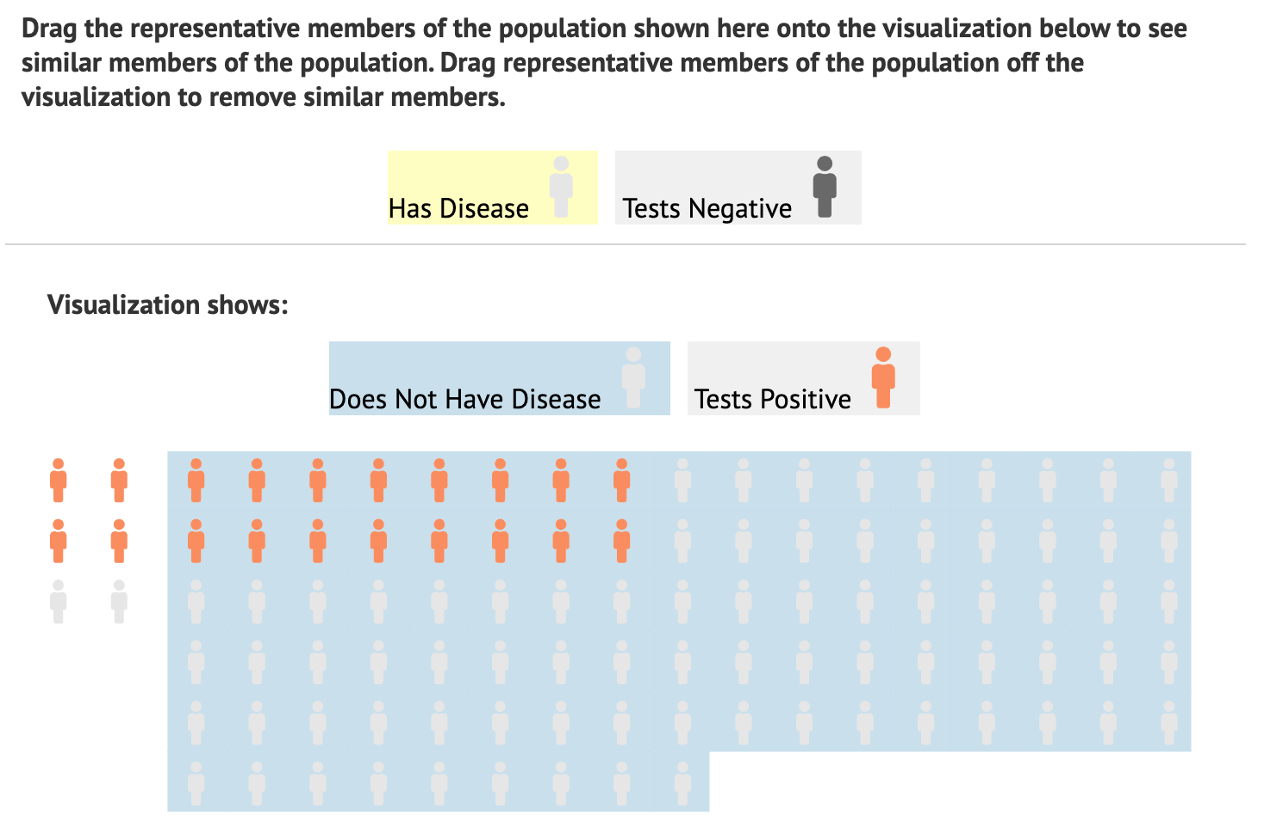}  
    & \includegraphics[width=.29\linewidth]{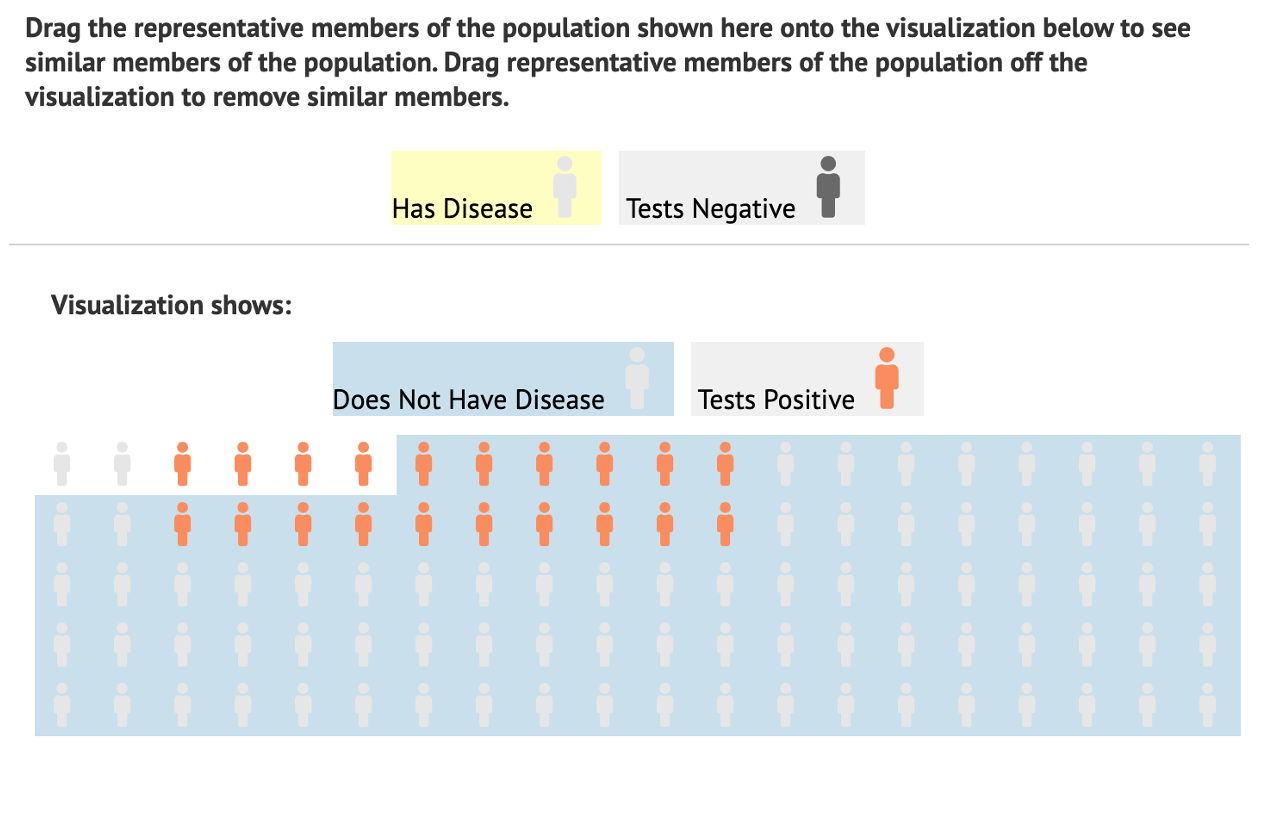}  
    & \includegraphics[width=.29\linewidth]{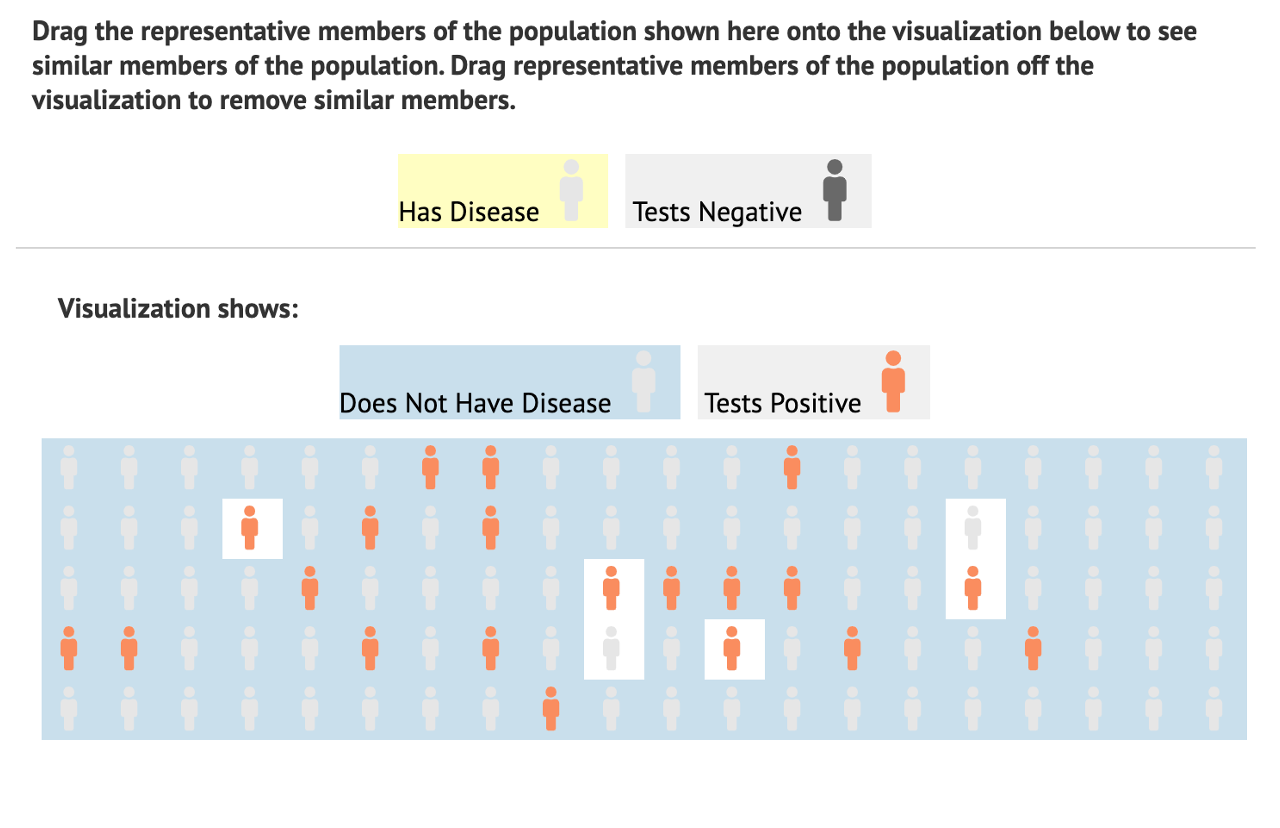}
    \\
    & \rotatebox{90}{\textit{hover}} 
    & \includegraphics[width=.29\linewidth]{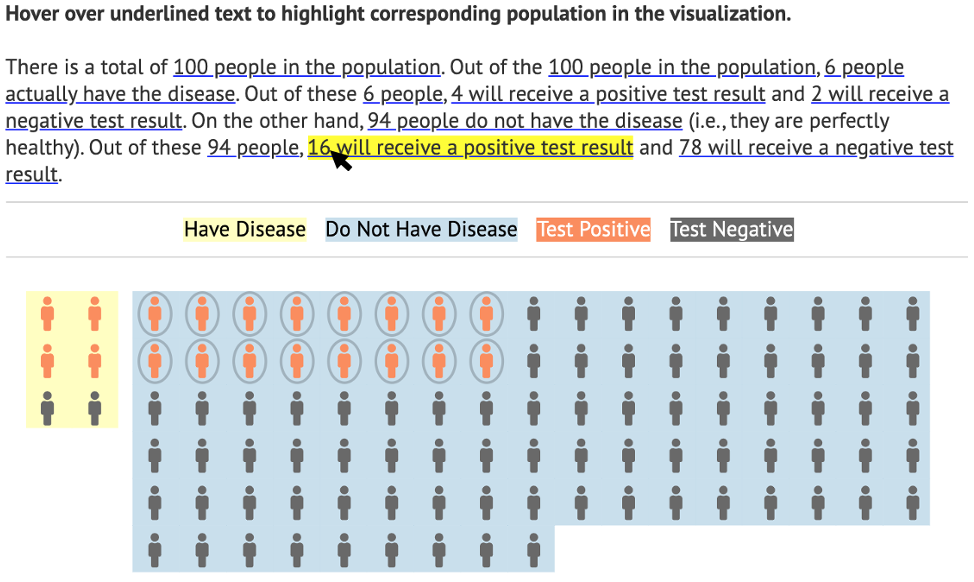}  
    & \includegraphics[width=.29\linewidth]{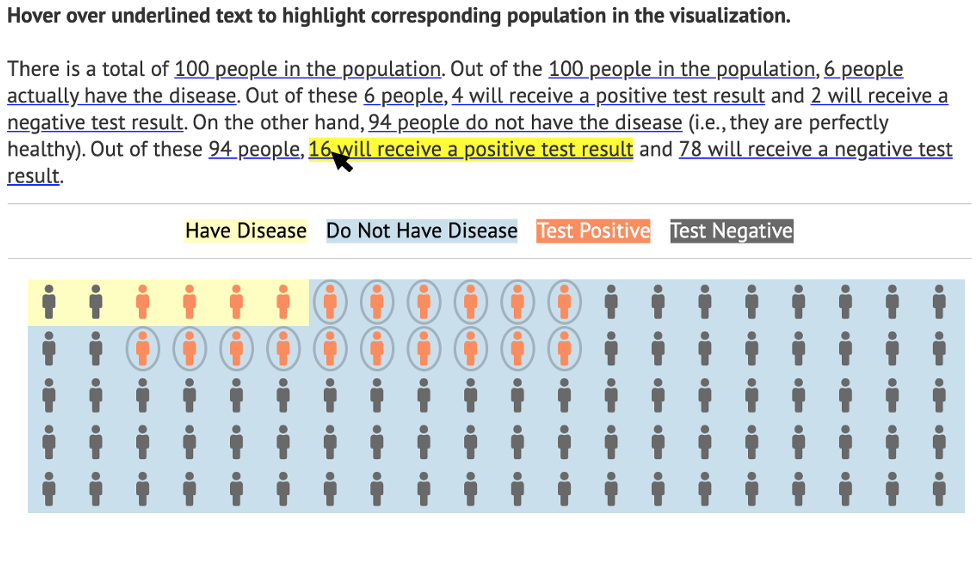}  
    & \includegraphics[width=.29\linewidth]{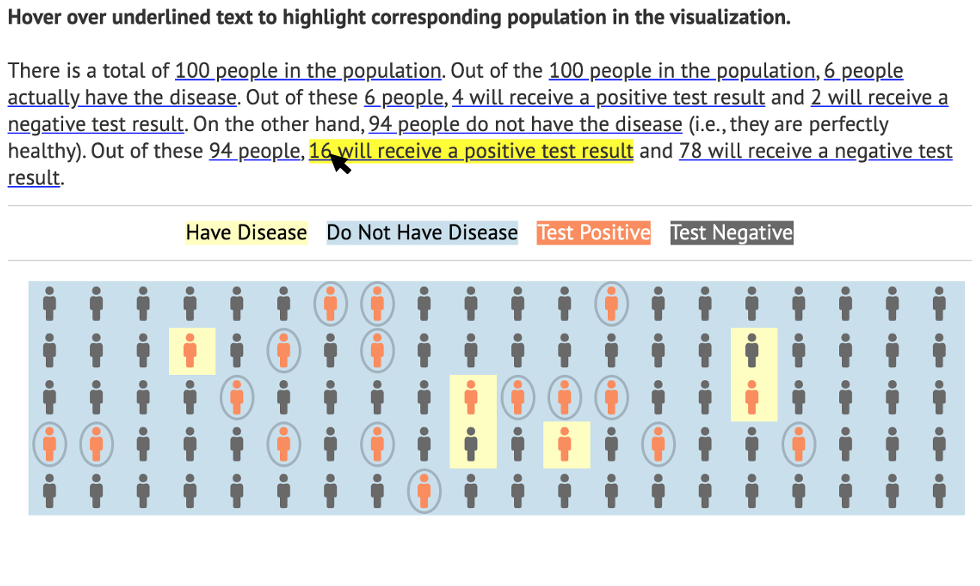}
    \\
    & \rotatebox{90}{\textit{tooltip}} 
    & \includegraphics[width=.29\linewidth]{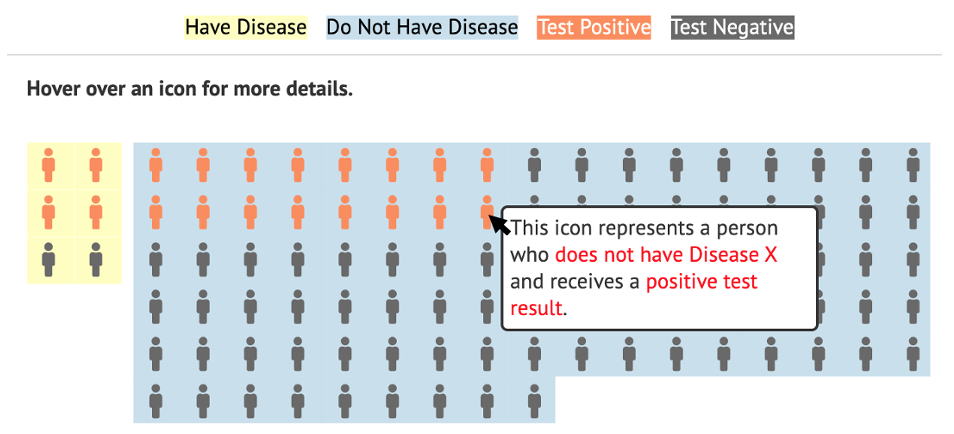}  
    & \includegraphics[width=.29\linewidth]{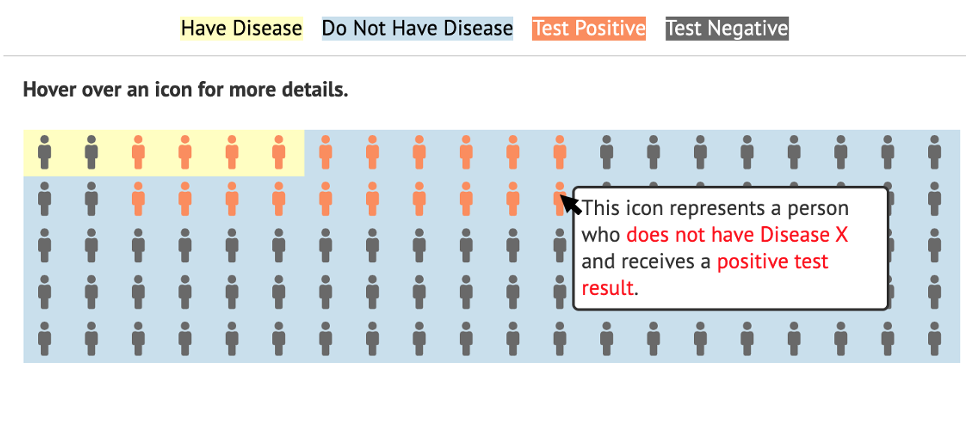}  
    & \includegraphics[width=.29\linewidth]{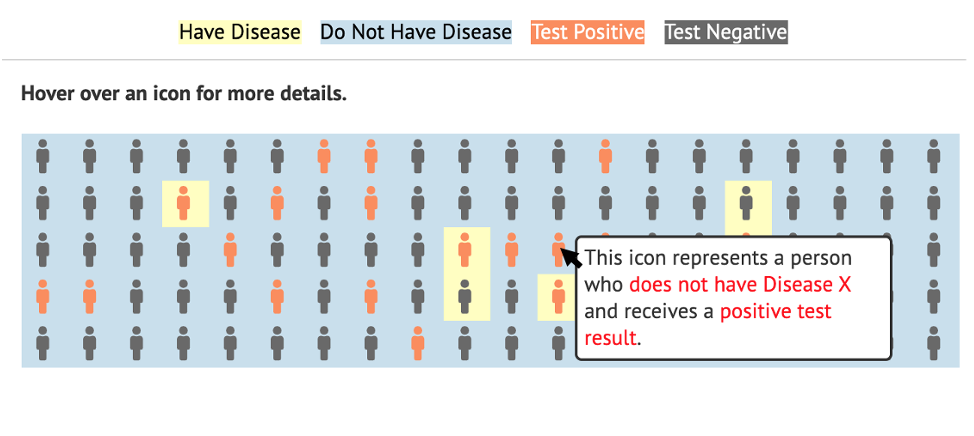} \\ \bottomrule
\end{tabular}
\end{center}
\label{tab:tabFullVis}
\Description[Examples of stimuli for Exp. 2]{Examples of each base visualization design and the cbAll / cbNone interaction (top row), drag interaction (second row), hover interaction (third row), and tooltip interaction (fourth row). In all cases the visualizations show an example of how the Test Positive and Do Not Have Disease populations would be highlighted or displayed.}
\end{table*}

\subsection{Visualization Designs}
Consistent with Experiment 1, each stimulus is an icon array that encodes the four key sub-populations in the Bayesian reasoning problem. 
Examples of stimuli are shown in Table \ref{tab:tabFullVis}. Below, we describe each experimental factor:

\begin{itemize}
    \item \textbf{Base Visualization}: \{ \textit{grouped, aligned, randomized} \}
    \item \textbf{Interaction Technique}: \{ \textit{checkboxes, drag and drop, hover, tooltips} \}
\end{itemize}

\subsubsection{\textbf{Base Visualizations}}
We use the same three \textit{icon array} designs as Experiment 1 (\textit{grouped}, \textit{aligned}, \textit{randomized}). 

\subsubsection{\textbf{Interaction Techniques}}
We test five different interaction techniques. We chose three of them (two types of checkboxes, and drag and drop) for consistency with prior work~\cite{tsai2011Interactive, khan2018Interactive}, and two additional techniques representative of the well known ``\textit{overview first, zoom and filter, details on demand}" mantra of visualization design~\cite{shneiderman1996Eyes}. We chose hover as an interaction representative of \textit{zoom and filter},
and tooltips as a representative of \textit{details on demand}. 

We recognize that even within the constraints of these five interaction techniques there are countless ways to design an interactive Bayesian reasoning visualization. Our goal in designing each interaction (similar to Experiment 1) was to enable participants to remove information irrelevant to answering the Bayesian reasoning question from the visualization, or to allow participants to more easily integrate information shown in the visualization and text. Below, we describe the implementation and design goals of each interaction technique:    

\begin{itemize} 
\item[] \textbf{\textit{Checkbox All}}: We included checkboxes in Experiment 2 for continuity with Experiment 1, and consistency with prior work~\cite{tsai2011Interactive}. We modify the interaction from Experiment 1 such that any sub-population ``checked" on the legend is shown on the visualization with color, and any sub-population ``unchecked" on the legend is shown on the visualization with light grey placeholders. This enabled participants to check off only populations explicitly mentioned in the Bayesian reasoning question (\textsc{Have Disease} and \textsc{Test Positive}) in order to answer it. In contrast, the implementation of checkboxes in Experiment 1 required participants to check off \textsc{Have Disease}, \textsc{Do Not Have Disease}, and \textsc{Test Positive} in order to answer the Bayesian reasoning question.    

\item[] \textbf{\textit{Checkbox None}}: Checkbox None (\textit{cbNone}) is identical to \textit{cbAll}, except all checkboxes are unchecked by default. In other words, the page loads with only light grey placeholders shown on the visualization. 
   
\item[] \textbf{\textit{Drag and Drop}}: Drag and drop (\textit{drag}) is a direct manipulation interaction. It was chosen to be consistent with interaction tested in prior work~\cite{khan2018Interactive}. We designed \textit{drag} with the intent of providing the same direct encoding benefits as checkboxes. In practice, \textit{drag} functions identically to \textit{cbNone} except that participants drag legend labels onto and off of the visualization.

\item[] \textbf{\textit{Hover}}: Hover (\textit{hover}) is a filter interaction. Prior studies suggest that users struggle to integrate text and visualization when performing Bayesian reasoning~\cite{ottley2016Bayesian, micallef2012Assessing, ottley2019Curious}. We design \textit{hover} with intent to help users overcome this hurdle by drawing a clearer connection between text and visualization. As participants hover their mouse over areas of text describing sub-populations in the visualization, the text and corresponding sub-population are highlighted. 

\item[] \textbf{\textit{Tooltip}}: Tooltips (\textit{tooltip}) are an example of details on demand. Similar to \textit{hover}, we design this interaction with the intent to help users integrate text and visualization more easily. To facilitate this, we reduce the distance users need to move their eyes to integrate the text and visualization by directly overlaying the two. When a participant hovers their mouse over any icon in the visualization a text box appears describing to which of the four sub-populations that particular icon belongs.  

\end{itemize}

It is relevant to note that although these techniques differ, all interaction techniques provide the same amount of information as the static visualization condition. The use of interactions does not add more information. Since Bayesian reasoning is the manipulation of 4 basic values (true positive, true negative, false positive, and false negative), the goal of the interaction techniques is to help the user in two ways: (1) isolating values of interest in the visualization, and (2) drawing connections between the visual representation and the textual description of the problem.

\subsection{Task}
We run a between-subjects 5 \{\textit{interaction techniques}\} \textsc{x} 3 \{\textit{base visualizations}\} factor experiment. The same textual description and questions as in Experiment 1 (Section \ref{sec:questions}) are used for this experiment. 

    
    \begin{table}[ht]
     \caption{Sample size (N) for each condition in Experiment 2.}
    \resizebox{0.4\textwidth}{!}{
    \centering
    \begin{tabular}{ccccc} \toprule
        & \textit{grouped} & \textit{aligned} & \textit{randomized} &  Total\\ \midrule
    \textit{cbAll}   & 119     & 129    & 123   & 371    \\ 
    \textit{cbNone}  & 147     & 125    & 122   & 394  \\ 
    \textit{drag}    & 102     & 115    & 129   & 346  \\ 
    \textit{hover}   & 131     & 136    & 167   & 434   \\ 
    \textit{tooltip} & 147     & 154    & 134   & 435 \\ 
    Total      & 646      & 659    & 675   & 1980 \\ \bottomrule
    \end{tabular}}
    \label{tab:fullN}
    \end{table}
    
  
    \begin{table}[ht]
    \caption{Experiment 2 participant demographics. }
    \resizebox{0.45\textwidth}{!}{
    \centering
    \begin{tabular}{ll} \toprule
      \textit{Measure} & \textit{Percentages} \\ \midrule
    N                                                                                               & 1,980                                                                                                                                                 \\ 
    Age                                                                                             & \begin{tabular}[c]{@{}l@{}}18-24: 7.8\%, 25-39: 50.8\%, 40-49: 20.4\%,\\ 50-59: 12.5\%, 60+: 8.2\%\end{tabular}                                       \\ 
    Gender                                                                                          & \begin{tabular}[c]{@{}l@{}}Female: 54.1\%, Male: 45.1\%,\\ Non-Binary: 0.5\%\end{tabular}                                                             \\ 
    Education                                                                                       & \begin{tabular}[c]{@{}l@{}}High School: 27.1\%, Bachelors: 49.8\%,\\ Masters: 15.1\%, PhD: 1.5\%, Other: 6.0\%\end{tabular}                           \\ 
    \begin{tabular}[c]{@{}l@{}}Expertise with \\ Statistical \\ Visualization\end{tabular}          &             \begin{tabular}[c]{@{}l@{}}Novice: 15.2\%, Low-intermediate: 21.4\%, \\ Intermediate: 38.8\%,\\ High-intermediate: 18.5\%, Expert: 5.4\%\end{tabular} \\ 
    \begin{tabular}[c]{@{}l@{}}Statistical Training \\ 1(none) - \\ 5 (highly trained)\end{tabular} & \begin{tabular}[c]{@{}l@{}}1: 29.6\%, 2: 22.3\%, 3: 21.8\%,\\ 4: 16.9\%, 5: 8.0\%\end{tabular}                                                        \\ \bottomrule
    \end{tabular}}
    \label{tab:fullDemo}
    \end{table}
    

\subsection{Participants}
We recruited 2,149 participants from Amazon Mechanical Turk. Participation was restricted to workers in the United States with an approval rating greater than $90$ percent. Participants were paid a base rate of $\$0.80$ for participation, plus a bonus of $\$0.10$ for every correct answer. 

Participants who skipped entire sections of the experiment or did not follow instructions ($N = 35$), and participants who self-identified as colorblind ($N = 134$) were dropped from the data set. 
This left $N = 1,980$ participants distributed among stimuli as shown in Table \ref{tab:fullN}. Demographics of participants are shown in Table~\ref{tab:fullDemo}.  

\subsection{Procedure}
Experiment 2 follows the same procedure as Experiment 1 with two exceptions. First, there is no instruction page, instead instructions are provided alongside each stimulus. And second, participants are asked to complete a an additional NASA-TLX~\cite{NASATLX} survey after the completion of the main study to measure task difficulty.

\subsection{Research Questions}
We analyze data from Experiment 2 to answer these questions: 

\begin{itemize} 
	
	\item[] \textbf{Q2.1} \textbf{Do different interaction techniques have different effects on accuracy in Bayesian reasoning?}
	We test whether participants who saw any interactive visualization will be more accurate in answering the Bayesian reasoning question than participants who saw a static visualization. 
	
	\item[] \textbf{Q2.2} \textbf{Is the effect of interaction moderated by interaction design and the underlying static visualization design?}
	We test if differences in accuracy will be modulated by base (\textit{grouped, aligned, randomized}) visualization design, and interaction technique (\textit{cbAll, cbNone, drag, hover, tooltip}).
	
	\item[] \textbf{Q2.3} \textbf{How does the effect of different interaction techniques change given different spatial abilities?}
	We split participants into high and low spatial ability groups and repeat the analyses for \textbf{Q2.1} within each group.

	\item[] \textbf{Q2.4} \textbf{Does  underlying static  visualization  design moderate  the  effect of  interaction  techniques within spatial ability groups?}
	We split participants into high and low spatial ability groups and repeat the analyses for \textbf{Q2.2} within each group.
\end{itemize}

\subsection{Findings}
As in Experiment 1, participants' answers were considered correct only if they answered both parts of the two-part question correctly. In order to compare our interactive visualizations against a static one, we included the \textit{static} group from Experiment 1 as an ``interaction technique" in our analysis (this brings the total number of participants to $N = 2,206$). From here forward the term ``interaction techniques" refers to \textit{cbAll}, \textit{cbNone}, \textit{drag}, \textit{hover}, \textit{tooltip}, and \textit{static}. Our analysis script is included in supplemental materials, however under the guidelines of our IRB we cannot release data. 

    
    \begin{figure}[ht]
    \includegraphics[width=0.45\textwidth]{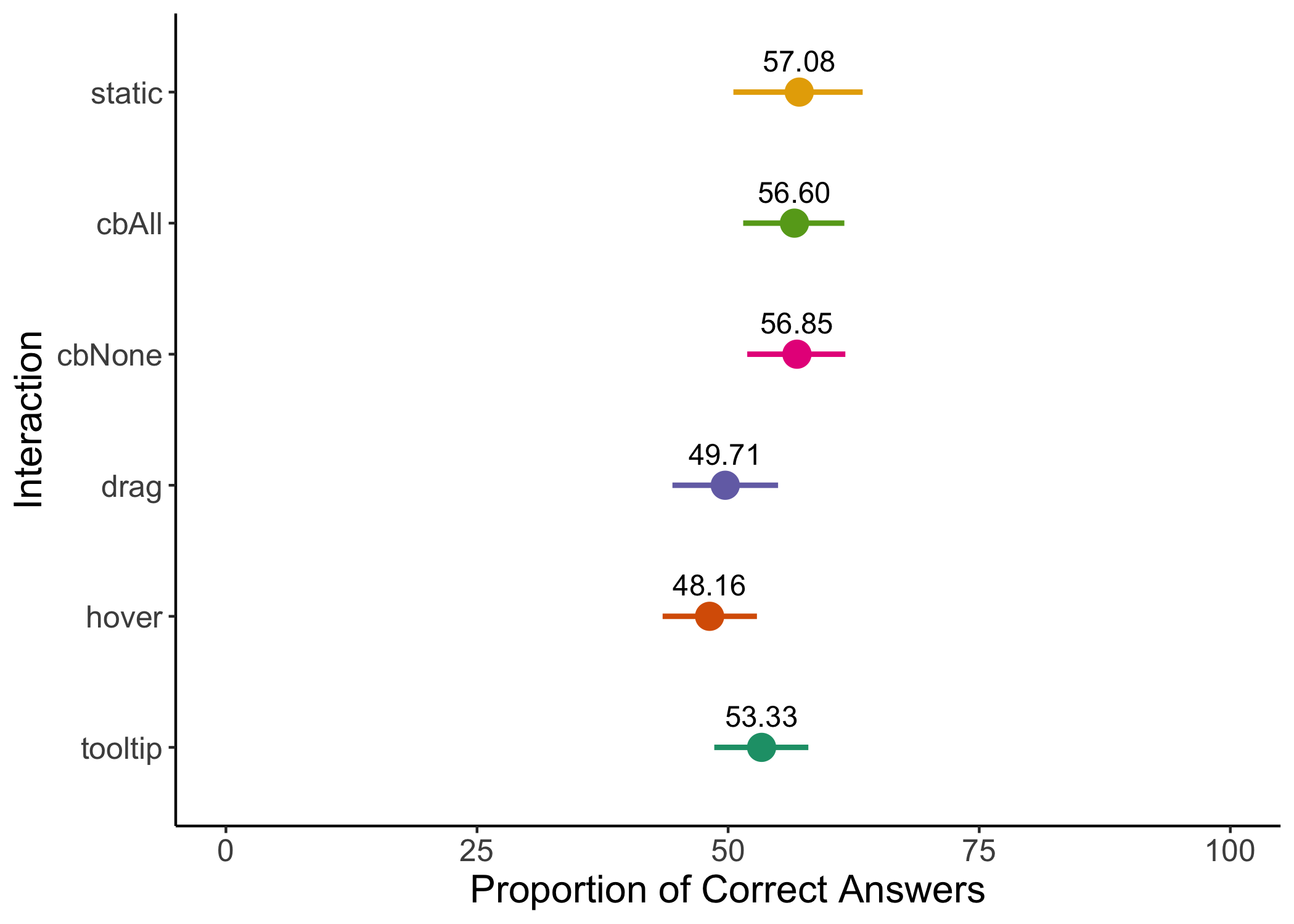}
    \caption{Portion of participants answering the Bayesian reasoning task correctly given each interaction technique. Bars represent a 95\% logit transformed confidence interval.}
    \label{fig:exp2_interactions}
     \Description[All interactions and static perform similarly.]{Dot chart showing proportions of correct answers for participants assigned to each interactive and static visualization. 95 percent logit transformed confidence intervals are also shown. Proportions are about equal in all cases, and confidence intervals overlap significantly.}
    \end{figure}
    
  
    \begin{figure}[ht]
    \includegraphics[width=0.45\textwidth]{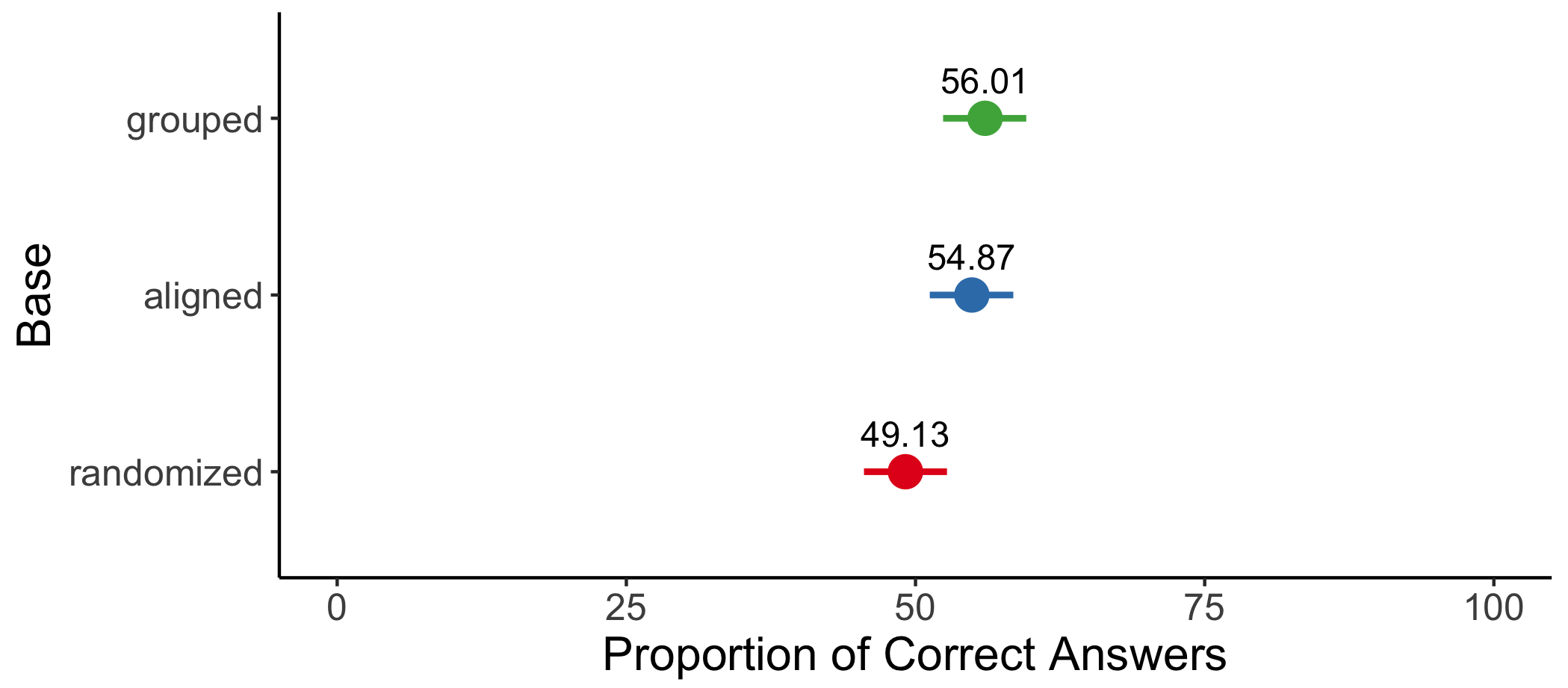}
    \caption{Portion of participants answering the Bayesian reasoning task correctly by base visualization. Bars represent a 95\% logit transformed confidence interval}
    \label{fig:exp2_bases}
     \Description[Performance is worse with randomized.]{Dot chart showing proportions of correct answers for participants assigned to the each base visualization. 95 percent logit transformed confidence intervals are also shown. There is a significant decrease in accuracy from the grouped to randomized base. No other comparisons show significantly different proportions of correct answers.}
    \end{figure}
    

\subsubsection{\textbf{Do different interaction techniques have different effects on accuracy in Bayesian reasoning?}}

We perform a 6-sample test for equality of proportions of $accuracy \sim interaction\_techniques$ with the null hypothesis that there is no difference in proportions of correct answers. We find no statistically significant difference in accuracy between participants using different interaction techniques ($\chi^2(5, N = 2206) = 11.33, p = 0.05$), and therefore fail to reject the null hypothesis. 
As shown in Figure \ref{fig:exp2_interactions}, 
we observe similar proportions of correct answers across interactions. 
This suggests that \textit{the design of an interactive technique does not affect the value-add of interaction}.

Similar to Experiment 1, out of the $371$ participants assigned to \textit{cbAll}, only $44\%$ actually interacted.
This proportion was significantly higher for all other interactive conditions. For each interaction technique we perform a 2-sample test for equality of proportions of $accuracy \sim interacted\_or\_not$ with the null hypothesis that there is no difference in proportions of correct answers. Interestingly \textit{drag} is the only condition in which participants who interacted were significantly more accurate than those who did not ($\chi^2(1, N = 346) = 18.49, p < 0.001$), and therefore is the only case in which we reject the null hypothesis. The 95\% confidence interval (using Wilson's score method) for the difference between  proportions is [-38.69, -14.93].  Further analysis is included in supplemental materials.

\subsubsection{\textbf{Is the effect of interaction moderated by interaction design and the underlying static visualization design?}}

We compare participants' accuracy across base visualizations with a 3-sample test for equality of proportions. We look at $accuracy \sim base\_visualization$ with the null hypothesis that there are no differences in proportions of correct answers. We find a significant difference in accuracy by base ($\chi^2(2, N = 2206) = 8.09, p = 0.02$), and therefore reject the null hypothesis. Pairwise comparisons with a Bonferroni corrected alpha ($0.02$) show a significant difference between the \textit{grouped} and \textit{randomized} bases ($\chi^2(1, N = 1477) = 6.74, p < 0.01$). The 95\% confidence interval (using Wilson's score method) for the difference between these two proportions is $[-12.10, -1.67]$. As shown in Figure \ref{fig:exp2_bases}, the proportion of correct answers in the \textit{grouped} base is higher than that for the \textit{randomized} base. This suggests that \textit{design of the underlying static visualization has an effect on participants' accuracy in a Bayesian reasoning task}.

\begin{figure}[ht]
    
    \includegraphics[width=0.45\textwidth]{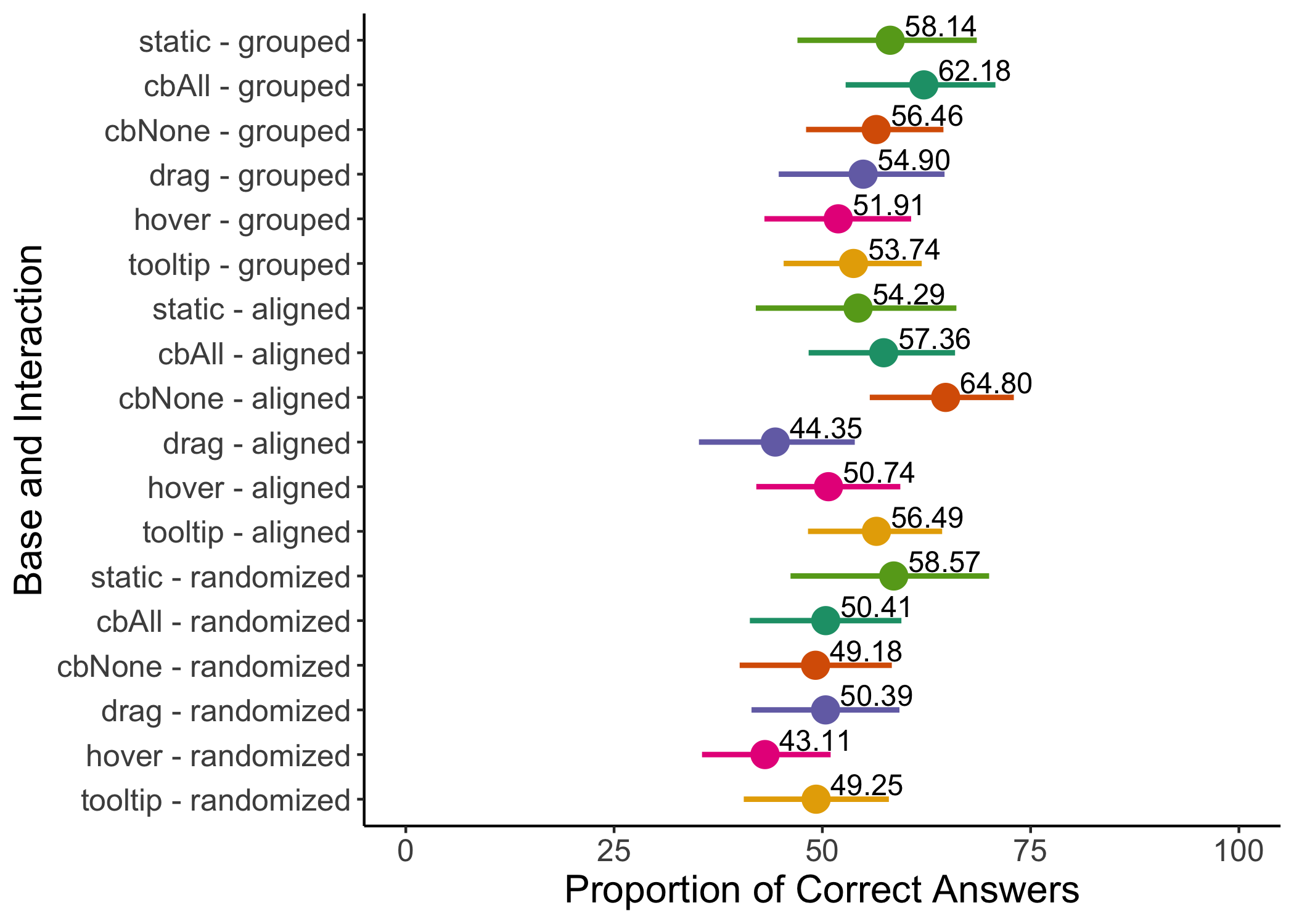}
    \caption{Portion of participants answering the Bayesian reasoning task correctly given each interaction technique and base visualization. Bars represent 95\% logit transformed confidence intervals.}
    \label{fig:exp2_interactions_by_base}
     \Description[All interactive and base visualization combinations perform similarly.]{Dot chart showing proportions of correct answers for participants assigned to each combination of interactive technique and base visualization. 95 percent logit transformed confidence intervals are also shown. Proportions are about equal in all cases, and confidence intervals overlap significantly.}
\end{figure}

To investigate whether there is an interaction effect between base visualization and interaction technique we perform an 18-sample test for equality of proportions. We look at $accuracy \sim \{base\_visualization\} X \{interaction\_technique\}$ with the null hypothesis that there are no differences in proportions of correct answers. We find no significant difference ($\chi^2(17, N = 2206) = 27.92, p = 0.05$) and therefore fail to reject the null hypothesis. 

Figure \ref{fig:exp2_interactions_by_base} shows proportions of participants answering the Bayesian reasoning task correctly given each combination of base visualization design and interaction technique. Similar to Experiment 1, we observe that any interactive technique added to the \textit{randomized} base resulted in lower accuracy on average than the static version of \textit{randomized}. This is not the case for the \textit{grouped} and \textit{aligned} bases. Practically speaking, this suggests that \textit{the combination of the \textit{randomized} base design with interaction may reduce average accuracy on the Bayesian reasoning task.}   

\subsubsection{\textbf{How does the effect of different interaction techniques change given different spatial abilities?}}
 
Prior work demonstrates that spatial ability is a significant predictor of performance on a Bayesian reasoning task and that people with low spatial ability in particular struggle to perform Bayesian inference~\cite{ottley2016Bayesian}. Based on this prior finding, we assume a difference in accuracy for participants with high versus low spatial ability and re-run our previous analyses stratified by spatial ability. We assess how people with high and low spatial ability react to the interaction techniques and base visualization designs we test.

\begin{figure}[ht]
    \includegraphics[width=0.45\textwidth]{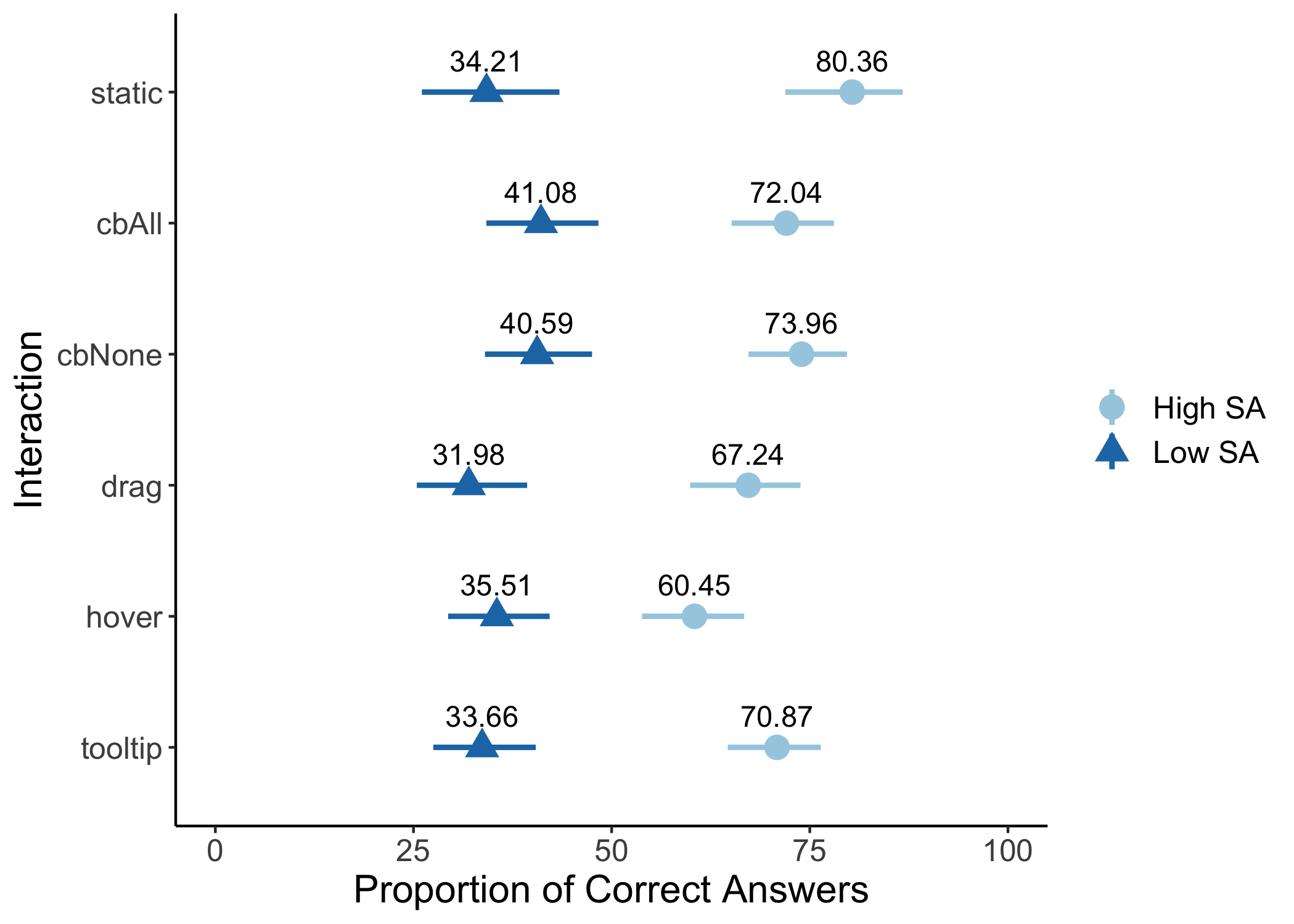}
    \caption{Portion of participants answering the Bayesian reasoning task correctly by spatial ability (SA) and interaction technique. Bars represent 95\% logit transformed confidence intervals.}
    \label{fig:exp2_sa_by_interaction}
     \Description[Static visualization outperforms hover for high SA.]{Dot chart showing proportions of correct answers for participants assigned to each interactive and static visualization. Proportions are shown for participants with high SA and low SA, and are about equal within each SA group in all cases, except for the static visualization vs the hover visualization for participants with high SA. Participants with high SA perform significantly worse with the hover visualization than with the static visualization. }
    \end{figure}

\begin{figure}[ht]
    \includegraphics[width=0.45\textwidth]{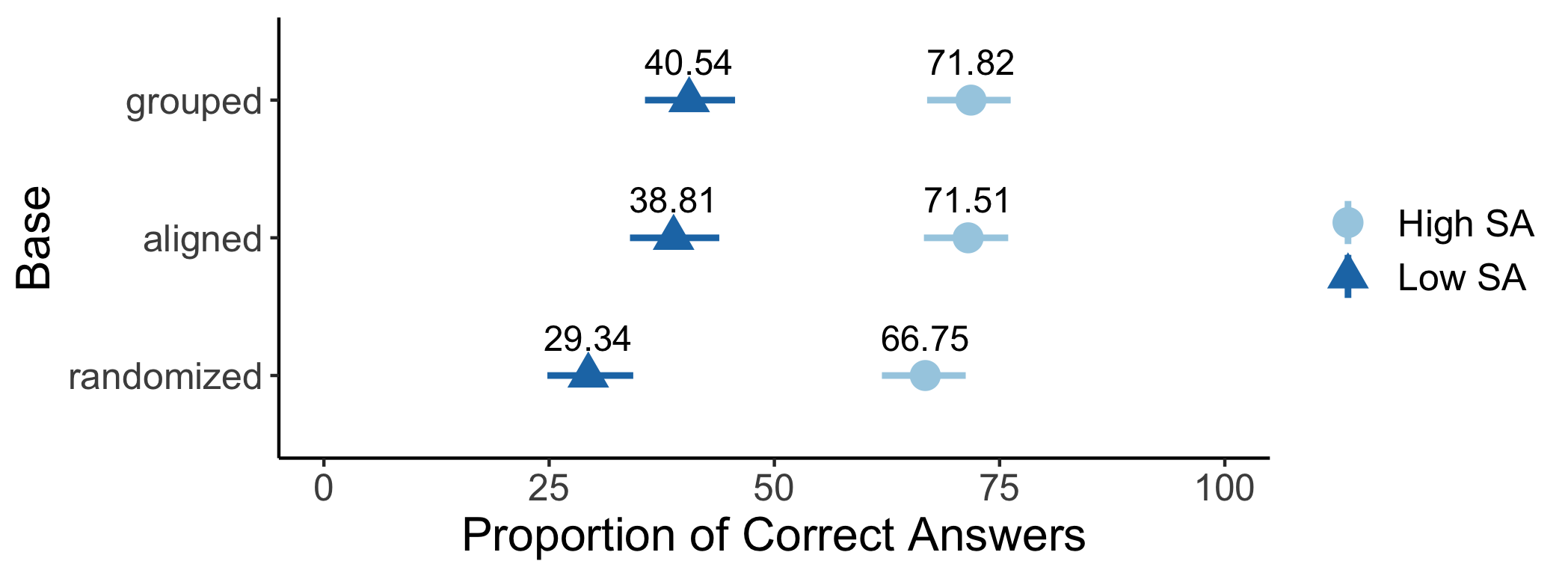}
    \caption{Proportion of participants answering the Bayesian reasoning task correctly by base visualization for each spatial ability group. Bars represent 95\% logit transformed confidence intervals.}
    \label{fig:exp2_bases_by_sa}
     \Description[Low SA participants perform worse with randomized.]{Dot chart showing proportions of correct answers for participants assigned to each base visualization. Proportions are shown for participants with high SA and low SA, and are about equal within the high SA group. Within the low SA group there is a significant decrease in performance given the randomized visualization compared to the grouped and aligned visualizations.}
    \end{figure}

Participants' spatial ability scores ranged from $-5$ to $20$. The median score was $6.25$. Consistent with prior work~\cite{ottley2016Bayesian}, we assigned participants with spatial ability scores greater than or equal to the median to the high spatial ability group, and participants with scores less than the median to the low spatial ability group. Figures \ref{fig:exp2_sa_by_interaction} and \ref{fig:exp2_bases_by_sa} plot performance on the Bayesian reasoning task by spatial ability group (high or low). In both of these figures we observe no overlaps in 95\% confidence intervals for proportions of correct answers of participants with high versus low spatial ability. This supports our assumption that there is a difference in accuracy for people with high versus low spatial ability. The following sections report analyses stratified by spatial ability group.

First, we perform a 6-sample test for equality of proportions of $accuracy \sim interaction\_technique$ with the null hypothesis that there are no differences in proportions of correct answers, stratified by spatial ability group. 

Within the high spatial ability group we find a statistically significant difference in accuracy between participants using different interaction techniques ($\chi^2(5, N =  1114) = 17.76, p < 0.005$), and therefore reject the null hypothesis. Pairwise tests with a Bonferroni corrected alpha ($0.003$) show a significant difference between the \textit{hover} and \textit{static} interactions ($\chi^2(1, N =  332) = 12.44, p < 0.001$). The 95\% confidence interval (using Wilson's score method) for the difference between these two proportions is $[9.43, 30.37]$. A larger proportion of participants with high spatial ability answered the Bayesian reasoning task correctly using the \textit{static} visualization versus the \textit{hover} visualization (Figure \ref{fig:exp2_sa_by_interaction}). 

Within the low spatial ability group we find no statistically significant difference in accuracy between participants using different interaction techniques ($\chi^2(5, N =  1092) = 5.72, p = 0.33$), and therefore fail to reject the null hypothesis.

These results suggest that \textit{for people with high spatial ability, an interactive \textit{hover} visualization can significantly decrease accuracy on Bayesian inference compared to a \textit{static} visualization.} And that \textit{for people with low spatial ability, adding interaction to a static visualization does not significantly effect reasoning accuracy}. 

   
\subsubsection{\textbf{Does underlying static visualization design moderate the effect of interaction techniques within spatial ability groups?}}

Within each spatial ability group we compare participants' accuracy across bases by performing a 3-sample test for equality of proportions of $accuracy \sim base\_visualization$, with the null hypothesis that there are no differences in the proportions of correct answers. 

Within the high spatial ability group we find no significant difference in accuracy by base ($\chi^2(2, N = 1114) = 2.93, p = 0.23$), and therefore fail to reject the null hypothesis. 
Figure \ref{fig:exp2_bases_by_sa} shows near equal proportions of participants with high spatial ability answering the Bayesian reasoning questions correctly across base visualizations.

Within the low spatial ability group we find a significant difference in accuracy by base ($\chi^2(2, N = 1092) = 11.23, p < 0.005$), and therefore reject the null hypothesis. Pairwise tests with a Bonferroni corrected alpha ($0.02$) show participants with low spatial ability assigned the \textit{randomized} base performed significantly worse than those assigned the \textit{grouped} ($\chi^2(1, N =  721) = 9.43, p < 0.005$) and \textit{aligned} ($\chi^2(1, N =  722) = 6.77, p < 0.01$) bases. The 95\% confidence intervals (using Wilson's score method) for the differences in these proportions are $[-18.38, -4.01]$ and $[ -16.62,  -2.32]$, respectively (Figure \ref{fig:exp2_bases_by_sa}). This suggests that \textit{for people with high spatial ability, performance on a Bayesian reasoning task is not affected by design of the underlying static visualization, and for people with low spatial ability it is}. 

Finally, within each spatial ability group we check for an interaction effect between base visualization design and interaction technique. We perform an 18-sample test for equality of proportions of $accuracy \sim  \{base\_visualization\} X \{interaction\_technique\}$ with the null hypothesis that there are no differences in proportions of correct answers. 

Within the high spatial ability group we find no significant differences ($\chi^2(17, N = 1114) = 23.64, p = 0.13$), and therefore fail to reject the null hypothesis. 

Within the low spatial ability group we again find no significant differences ($\chi^2(17, N = 1092) = 23.41, p = 0.14$), and therefore fail to reject the null hypothesis. 

\begin{figure}[ht]
    \includegraphics[width=0.45\textwidth]{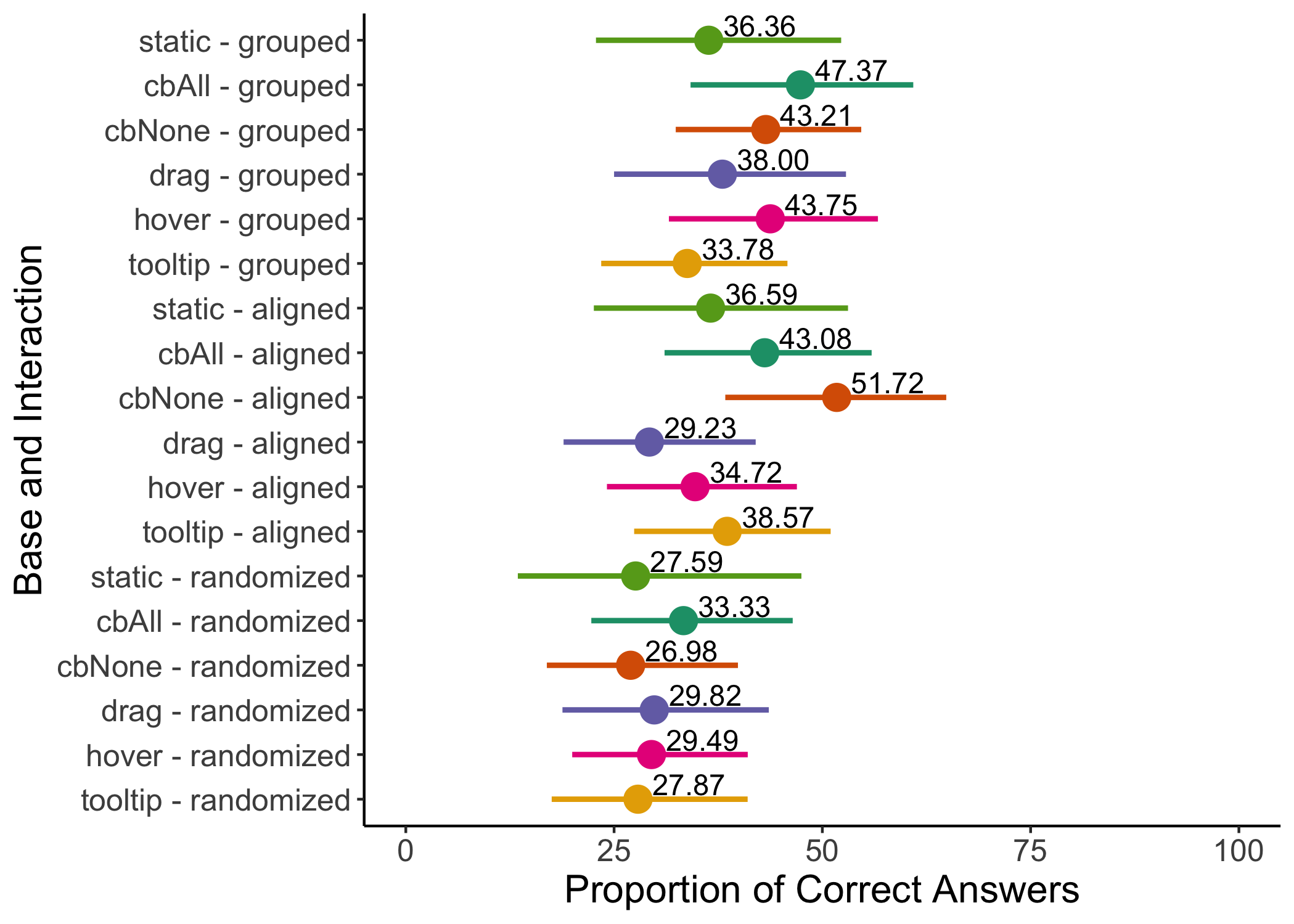}
    \caption{Portion of participants with \textbf{low spatial ability} answering the Bayesian reasoning task correctly given each interaction technique and base visualization. Bars represent a 95\% logit transformed confidence intervals.}
    \label{fig:exp2_corr_by_int_base_SA_low}
     \Description[Combinations of interaction technique and base visualization perform similarly for low SA.]{Dot chart showing proportions of correct answers for participants with low SA assigned to each combination of interaction technique and base visualization. Proportions are about equal in all cases (95 percent confidence intervals all overlap).}
    \end{figure}

\begin{figure}[ht]
    \includegraphics[width=0.45\textwidth]{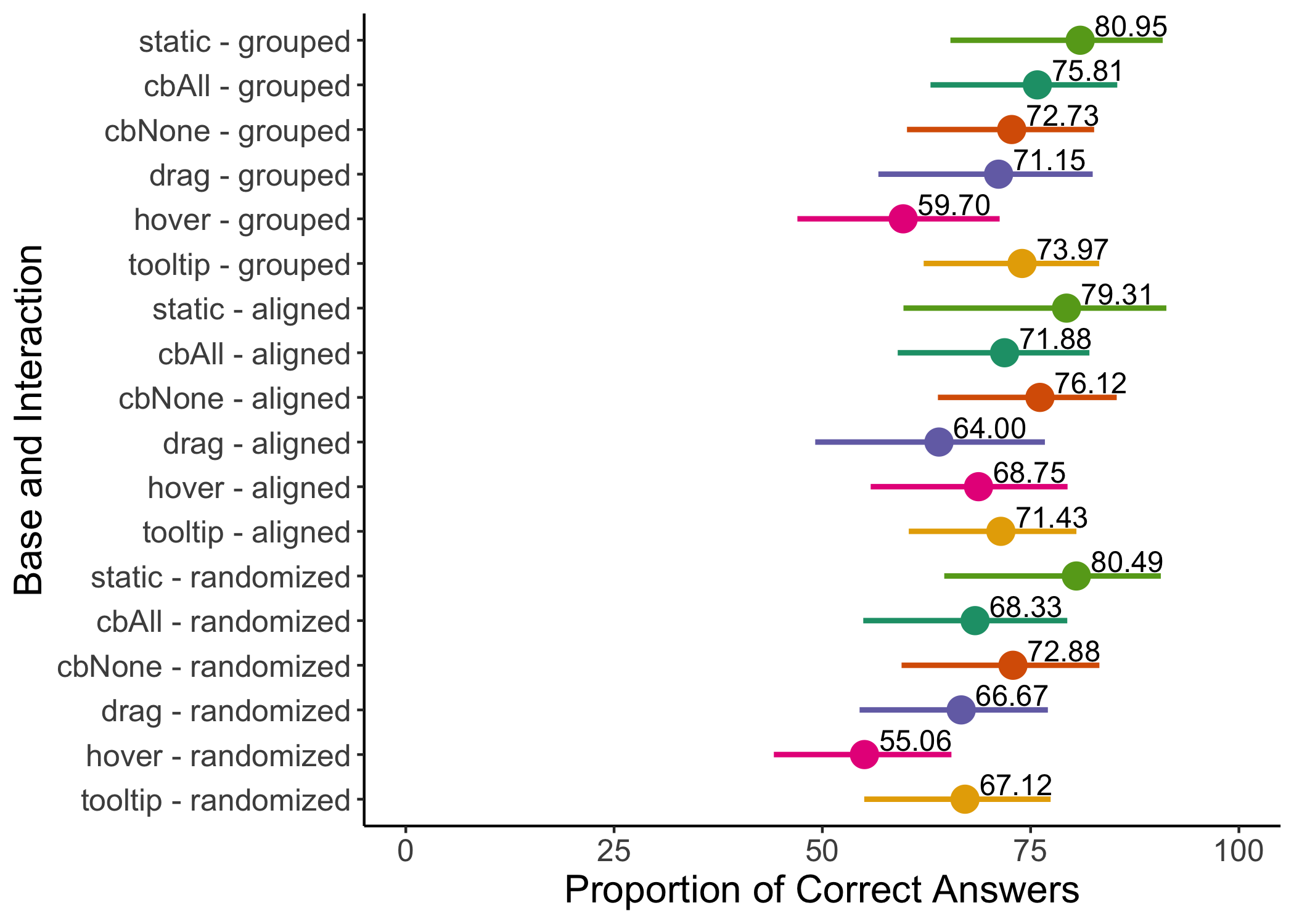}
    \caption{Portion of participants with \textbf{high spatial ability} answering the Bayesian reasoning task correctly given each interaction technique and base visualization. Bars represent a 95\% logit transformed confidence intervals.}
    \label{fig:exp2_corr_by_int_base_SA_high}
     \Description[Combinations of interaction technique and base visualization perform similarly for high SA.]{Dot chart showing proportions of correct answers for participants with high SA assigned to each combination of interaction technique and base visualization. Proportions are about equal in all cases (95 percent confidence intervals all overlap).}
    \end{figure}

As shown in Figures \ref{fig:exp2_corr_by_int_base_SA_low} and \ref{fig:exp2_corr_by_int_base_SA_high} we observe near equal proportions of correct answers in all combinations of base visualizations and interaction techniques. This suggests that \textit{for people with high and low spatial ability the value-add of interaction is not modulated by underlying static visualization design and interaction technique}.   


    
\section{Discussion} 
\label{sec:exp2_discussion}

Three of the interaction techniques tested in Experiment 2 are similar to those tested in prior work (\textit{cbAll, cbNone, drag})~\cite{tsai2011Interactive, khan2018Interactive}. Our findings from both experiments suggest that neither \textit{checkbox} interaction significantly improves performance on a Bayesian reasoning task compared to a \textit{static} visualization. While prior work has suggested that an interactive checkbox visualization can increase accuracy on a Bayesian reasoning task~\cite{tsai2011Interactive}, that work compared an interactive visualization coupled with a textual description of the problem using frequencies to a textual description of the problem using probabilities. Our findings add nuance to this work by explicitly comparing an interactive and static visualization, and by using a constant wording of the text. These differences in experimental set up likely explain the discrepancy between findings in prior work and our findings. Similarly, we find that the \textit{drag} interaction does not significantly change performance on the Bayesian reasoning task compared to a \textit{static} visualization. These findings are consistent with prior work~\cite{khan2018Interactive}, which found adding a drag and drop interaction to a double tree diagram decreased performance on a Bayesian reasoning task, but not significantly so.    

To broaden the scope of our results, we include interactions representative of the popular ``\textit{overview first, zoom and filer, details on demand}" mantra of visualization design~\cite{shneiderman1996Eyes} (\textit{hover, tooltip}). Again we find neither of these interaction techniques lead to significantly better performance on Bayesian reasoning than a \textit{static} visualization.


Moreover, we find participants with high spatial ability have significantly worse accuracy given the interactive \textit{hover} visualization versus \textit{static}. We postulate that this is another case where adding interaction to an already complex task causes cognitive overload. 
Ottley et al.~\cite{ottley2016Bayesian} showed that integrating text and visualization is especially difficult for people with high spatial ability. As \textit{hover} draws a very explicit link between text and visualization, it is likely an extremely cognitively taxing interaction for high spatial ability participants and thus led to cognitive overload and decreased performance. We encourage future work to further investigate the nuanced relationship between cognitive load, interaction, and spatial ability. 


The results of Experiment 2 coupled with those of Experiment 1 suggest a perhaps unexpected answer to the question ``does interaction improve Bayesian reasoning with visualization?''
Despite our best efforts, none of the interaction techniques we test significantly improve participants' performance on a Bayesian reasoning task compared to a static visualization. Moreover, we identify several scenarios that suggest interaction  \textit{decreases} participants' performance. 
In Experiment 1, we observe adding checkboxes where all the boxes are pre-checked (\textit{cbAll}) to the \textit{randomized} base visualization decreases performance of participants (Figure \ref{fig:exp1_static_vs_int_by_base}). This decrease is not statistically significant, but has practical implications. Similarly, in Experiment 2 we observe statistically insignificant decreases in accuracy across all interaction techniques added to the \textit{randomized} design (Figure \ref{fig:exp2_interactions_by_base}). In addition, in Experiment 2 we see that across visualization designs participants with high spatial ability perform significantly worse with an interactive \textit{hover} visualization than with a \textit{static} visualization. While all of these findings are not statistically significant, together they suggest that interaction may not be universally beneficial to static Bayesian reasoning visualizations, and that this topic warrants further investigation.   

Our findings suggest that a well-designed static visualization can be as (if not more) beneficial to solving complex reasoning tasks as an interactive visualization.
In cases where the user's interactions result in additional information being shown on the visualization (e.g. panning a map), the value of the interaction is undisputed.
However, general claims that interaction can improve reasoning, and offer cognitive support can be called into question given the results of our experiments.
We observe more than one scenario in which the use of an interactive visualization can be detrimental, and we were unable to show any cases in which use of an interactive visualization led to significant improvement in Bayesian inference.
Therefore, we echo the sentiment made by researchers such as Lam~\cite{lam2008Framework} and van Wijk~\cite{wijk2005Value}, and practitioners like the New York Times, and suggest a cautious use of interactivity in cases where it does not add additional information to a visualization.

\section{Limitations and Future Work}

We acknowledge there are limitations to our work and that there remain open questions for future research.

Studies such as the ones presented in this paper are fundamentally limited in scope. While we made our best effort to select the most appropriate visualization designs and implement the best interactions based on existing literature, we acknowledge that there are infinite options that could be tested. It is plausible that there exists a combination of visualization and interaction techniques that can improve participants' abilities to solve the Bayesian reasoning task. 
However, our recommendation of a cautious use of interactivity as a reasoning aid remains true. 
In everyday designs of interactive visualizations, practitioners are unlikely to be able to carefully evaluate a large number of combinations of visualization and interaction designs. 
Our results show that, in those cases, practitioners should be cautious in adding interactions to a static visualization designed to help users perform reasoning tasks.

It is important to note that the interaction techniques used in our studies do not add new information to the static visualizations.
However, doing so would not aid in Bayesian reasoning. 
Adding new information is not meaningful because the Bayesian reasoning is inherently based on understanding relationships between four integer values (true positive, false positive, true negative, and false negative counts).
Moreover
showing the user the numerical answer to the Bayesian reasoning problem 
does not help them better understand the reasoning process behind that number. In practice, that reasoning process, more so than a number value, is critical to decision making (e.g. in a medical decision-making scenario~\cite{trevena2013presenting, han2011Representing}).  


In addition, we acknowledge that there are many different formulations of the Bayesian reasoning problem with different levels of sensitivity, specificity and disease prevalence and that perturbing these values could result in different findings than what we have presented here. Moreover, there are numerous static visualization designs with which we could have performed this study.   
However, we see this work as a starting point to a principled investigation of the costs and benefits of interaction. There are countless factors that could be manipulated and tested. In performing this study we chose to bound the static visualizations tested to one category (icon arrays) as well as the formulation of the Bayesian reasoning problem. Both of these choices were made in an effort to keep as much consistency as possible with prior work in this space. Icon arrays are one of the most popular and well studied visualizations in this context~\cite{micallef2012Assessing, ottley2019Curious, ottley2016Bayesian, brase2009Pictorial}, and the specific formulation of the Bayesian reasoning problem used in this work has been used in a number of other studies~\cite{khan2015Benefits, micallef2012Assessing, ottley2016Bayesian, tsai2011Interactive}. In future work we plan to explore the effects of different formulations of the Bayesian reasoning problem, as well as different static visualization designs.    

Finally, our results suggest interaction may impede performance on high cognitive load tasks. 
However, to the best of our knowledge, beyond theoretical guidelines (such as the one by Lam~\cite{lam2008Framework}), there is no empirical work on evaluating interaction techniques based on their effect on a user's cognitive load.
As a future work, we aim to investigate interaction techniques using a cognitive-load theory. The result of which we hope will provide a theoretical understanding on the outcome of the studies presented in this paper.
\section{Conclusion} 
This paper empirically shows how different interaction techniques and visualization designs affect users in solving Bayesian reasoning tasks. 
Through two crowdsourced studies, we evaluated five interaction techniques across three different static visualization designs. 
The results illustrate that the effect of interaction is largely dependent on the design of the underlying static visualization, and implementation of the interaction itself. Additionally, we observe that people with different spatial abilities react to interaction differently. 
These findings suggest that adding interaction to a static Bayesian reasoning visualization may not be beneficial, and in some cases can be detrimental. 
For example, we find adding interaction to certain designs of static Bayesian reasoning visualizations can decrease users' accuracy on Bayesian inference. 
Similarly, we find when people with high spatial ability use a hover Bayesian reasoning visualization, they perform a Bayesian reasoning task with significantly worse accuracy than they do with a static visualization. Based on these findings we conclude that interaction may not be as unanimously beneficial as it is often believed to be; in some cases a well designed static visualization can be as, if not more, effective.

%
\begin{acks}
This work was supported by grants from the Walmart Foundation (OAC-1940175, OAC-1939945, IIS-1452977, DGE-1855886), DARPA D3M (FA8750-17-2-0107), and NSF Grant No. 1755734. The authors would also like to thank Rob Jacob and Megan Monroe for their input. 
\end{acks}

\bibliographystyle{ACM-Reference-Format}
\bibliography{main}



\end{document}